\documentclass[manuscript]{aastex} 
\usepackage{emulateapj5}
\usepackage{natbib} 
\usepackage{epsfig} 
\shortauthors{von Braun et al.} \shorttitle{EXPLORE/OC}
\begin{document}

%---------------------------------------------------------------------------

\title{Searching for Planetary Transits in Galactic Open Clusters: EXPLORE O/C}

%---------------------------------------------------------------------------

\author{Kaspar von Braun}
\affil{Carnegie Institution, Dept of Terrestrial Magnetism, 
	5241 Broad Branch Rd. NW, Washington, DC 20015}
\email{kaspar@dtm.ciw.edu}

\author{Brian L. Lee}
\affil{Dept of Astronomy and Astrophysics, University of Toronto, 
	60 St. George St., Toronto, Canada M5S 3H8}
\email{blee@astro.utoronto.ca}

\author{S. Seager}
\affil{Carnegie Institution, Dept of Terrestrial Magnetism, 
	5241 Broad Branch Rd. NW, Washington, DC 20015}
\email{seager@dtm.ciw.edu}

\author{H. K. C. Yee}
\affil{Dept of Astronomy and Astrophysics, University of Toronto,
	60 St. George St., Toronto, Canada M5S 3H8}
\email{hyee@astro.utoronto.ca}

\author{Gabriela Mall\'{e}n-Ornelas}
\affil{Harvard-Smithsonian Center for Astrophysics, 60 Garden Street, MS-15,
	Cambridge, MA 02138}
\email{gmalleno@cfa.harvard.edu}

\and{}
\author{Michael D. Gladders} 
\affil{Carnegie Observatories, 813 Santa Barbara St., Pasadena, CA 91101} 
\email{gladders@ociw.edu}

%---------------------------------------------------------------------------

\begin{abstract}

Open clusters potentially provide an ideal environment for the search
for transiting extrasolar planets since they feature a relatively
large number of stars of the same known age and metallicity at the
same distance. With this motivation, over a dozen open clusters are
now being monitored by four different groups.  We review the
motivations and challenges for open cluster transit surveys for
short-period giant planets. Our photometric monitoring survey
(EXPLORE/OC) of Galactic southern open clusters was designed with the
goals of maximizing the chance of finding and characterizing planets,
and of providing for a statistically valuable astrophysical result in
the case of no detections.  We use the EXPLORE/OC data from two open
clusters NGC 2660 and NGC 6208 to illustrate some of the largely
unrecognized issues facing open cluster surveys including severe
contamination by Galactic field stars ($>$ 80\%) and relatively low
number of cluster members for which high precision photometry can be
obtained.  We discuss how a careful selection of open cluster targets
under a wide range of criteria such as cluster richness,
observability, distance, and age can meet the challenges, maximizing
chances to detect planet transits. In addition, we present the
EXPLORE/OC observing strategy to optimize planet detection which
includes high-cadence observing and continuously observing individual
clusters rather than alternating between targets.
%and a new idea for dynamic observing to
%switch clusters after 10 nights if no full single transits are
%detected, given a 70\% probability to detect an existing single full
%transits in 10 consecutive nights.

\end{abstract}

%---------------------------------------------------------------------------


%---------------------------------------------------------------------------

\section{Introduction}\label{transits}

The EXPLORE Project \citep{msy03,yms03} is one of currently about 20
ongoing surveys\footnote{See
\url{http://star-www.st-and.ac.uk/$\sim$kdh1/transits/table.html},
maintained by K. Horne.} with the aim to detect transiting close-in
extrasolar giant planets (CEGPs; also referred to as `51-Peg type' or
`hot Jupiters', i.e., planets with radius of order Jupiter-radius,
orbital period of one to a few days, and transit durations of a few
hours) around Galactic main-sequence stars. Transit studies explore a
different parameter space in the search for extrasolar planets from
the very successful radial-velocity or ``wobble'' method. This method
based on detecting planets via the radial motions of their parent star
caused by the star's motion about the common center of mass \citep[see
for example table 3 in][]{bmv02}.  Fainter (and thus more) stars can
be monitored photometrically than spectroscopically.  Thus, more
distant environments can be probed for the existence of extrasolar
planets.
%not necessarily true, since 1-2m
%dedicated can be used for rv
%, using smaller telescopes.

All transiting planets have a measured radius, based on transit depth
and stellar radius. Knowledge of the planet's radius and mass plays an
important role in modeling internal structure of planets and hence the
formation, evolution, and migration of planetary systems \citep[see
for example][and references therein]{bgh00,gs02,bcb03}.

Transiting planets are currently the only planets whose physical
characteristics can be measured. In addition to mass and radius,
several parameters can be constrained from follow-up measurements.
For example, the fact that a transiting planet will be superimposed on
its parent star can be used to determine constituents of the planet's
atmosphere by means of transmission spectroscopy \citep{cbn02,vld03},
or put constraints on the existence of planetary moons or rings
\citep[e.g.,][]{bcg01}.  In addition, the secondary eclipse can
provide information about the planetary temperature or its emission
spectrum \citep{rds03, rdw03}.

At the time of writing, six transiting planets are known. HD209458b
\citep{cbl00,hmb00,bcg01} was discovered by radial-velocity
measurements \citep{hmb00, mdt00} and the transits were discovered by
photometric follow-up. OGLE-TR-56 \citep{upz02,uzs02,ktj03} was the
first planet discovered by the transit method, and the first of
currently four planets based on photometry of the OGLE-III survey
\citep{upz02,uzs02,ups03}: OGLE-TR-113 \citep{bps04,kts04},
OGLE-TR-132 \citep{bps04}, and OGLE-TR-111 \citep{pbq04}.  Very
recently, \citet{abt04} found a further transiting planet, TrES-1,
using telescopes with 10cm apertures. Over 20 transit searches are
currently ongoing to find more planets.

As part of the
EXPLORE\footnote{\url{http://www.ciw.edu/seager/EXPLORE/explore.htm}}
Project, we have recently begun a survey -- EXPLORE/OC -- of southern
open clusters (OCs) with the aim of detecting planetary transits
around cluster member stars. During the course of $\sim$3 years, we
hope to conduct searches of up to 10 OCs using the Las Campanas
Observatory (LCO) 1m Swope Telescope. To date we have monitored five
OCs (see \S \ref{targets_exp}).

In addition to EXPLORE/OC, at the time of writing there are currently
three OC planet-transit surveys underway\footnote{see also
\url{http://www.ciw.edu/kaspar/OC\_transits/OC\_transits.html}}:
\begin{itemize}
\item The Planets In Stellar Clusters Extensive Search
(PISCES\footnote{\url{http://cfa-www.harvard.edu/$\sim$bmochejska/PISCES}})
reported the discovery of 47 and 57 low-amplitude variables in the
open clusters NGC 6791 \citep{mss02} and NGC 2158 \citep{mss04},
respectively.
\item The University of St. Andrews Planet Search
(UStAPS\footnote{\url{http://crux.st-and.ac.uk/$\sim$kdh1/ustaps.html} and
\url{http://star-www.st-and.ac.uk/$\sim$dmb7}})
has monitored the OCs NGC 6819 \citep{shl02} and NGC 7789
\citep{bhb03} and published data on variable stars in NGC 6819
\citep{shl03}.
\item The Survey For Transiting Extrasolar Planets In Stellar Systems
(STEPSS\footnote{\url{http://www.astronomy.ohio-state.edu/$\sim$cjburke/STEPSS}}),
described in \citet{bdg03} and \citet{gbd02}. They have analyzed
monitoring data of the OC NGC 1245 (Burke et al., in preparation) and
determined its fundamental parameters \citep{bgd04}.  Analysis of
their data on NGC 2099 and M67 is currently ongoing.
\end{itemize}
The OC NGC 6791 was furthermore monitored for planetary
transits by \citet{bgt03}.  As all these surveyed OCs are located the
northern hemisphere,
EXPLORE/OC\footnote{\url{http://www.ciw.edu/seager/EXPLORE/open\_clusters\_survey.html}}
is currently the only OC survey operating in the south where most of
the Galactic OCs are located.

With the growing number of open star cluster surveys this publication
describes incentives, difficulties, and strategies for open cluster
planet transit surveys, thereby including a discussion on transit
surveys in general.  We use data from the first two targets (NGC 2660
and NGC 6208) from our program to illustrate the major issues for OC
transit surveys.

The concept and advantages of monitoring OCs for the existence of
transiting planets were originally described in \citet{j96}. Written
before the hot Jupiter planets were discovered, \citet{j96} focused on
12-year period orbits and long-term photometric precision required to
determine or put useful limits on the Jupiter-like planet
frequency. This paper is intended to be a modern version of
\citet{j96} based on the existence of short-period planets and
practical experience we have gained from both the EXPLORE and the
EXPLORE/OC planet transit surveys. Section \ref{transits_ocs}
motivates OC transit surveys. Section \ref{mainchallengeTS} addresses
the challenges facing transit surveys in general, and \S
\ref{mainchallengeOC} addresses challenges specifically facing OC
transit surveys. Section \ref{targets} focuses on strategies to select
OCs which are most suited for transit surveys and which minimize
challenges described in the previous sections.  The EXPLORE/OC
strategies concerning target selection, observing methods, photometric
data reduction, and spectroscopic follow-up observations are described
in \S \ref{targets_exp}, \S \ref{observing}, and \S \ref{photometry},
respectively. These Sections contain relevant preliminary results on
EXPLORE/OC's first two observed clusters NGC 2660 and NGC 6208. We
summarize and conclude in \S \ref{future}.

%---------------------------------------------------------------------------

\section{Motivation for Open Cluster Planet Transit Searches}
\label{transits_ocs}

Open clusters present themselves as ``laboratories'' within which the
effects of age, environment, and especially metallicity on planet
frequency may be examined.  Evidence that planet formation and
migration are correlated with metallicity comes from radial velocity
planet searches \citep{fv03}. The fact that no planetary transits were
discovered in the monitoring study of 47 Tuc by \citet{gbg00} may be
due to its low metallicity, or alternatively due to the high-density
environment in a system such as a globular cluster (or due to both
effects). The less-crowded OCs of the Milky Way offer a range of
metallicities and may thus be further used to disentangle the effects
of metallicity versus high-density environment upon planet frequency.

Monitoring OCs for the existence of planetary transits offers the
following incentives \citep[see also][]{j96, c03,lbm04,blm04}:

\begin{enumerate}
\item Metallicity, age, distance, and foreground reddening are either
known or may be determined for cluster members \citep[more easily and
accurately than for random field stars; see \S \ref{targets} and,
e.g., ][]{bgd04}. Thus, planets detected around open cluster stars
will immediately represent data points for any statistic correlating
planet frequency with age, stellar environment, or metallicity of the
parent star.
\item The planet-formation and planet-migration processes, and hence
planet frequencies, may differ between the OC, globular cluster, and
Galactic field populations.  Planet transit searches in OCs, together
with many ongoing transit field searches and GC surveys \citep[e.g.,
the ground-based work on 47 Tuc by][]{wsb03,wsb04}, enable comparison
between these different environments.
\item Specific masses and radii for cluster stars may be targeted
(within certain limits of other survey design choices, see \S
\ref{cadence}) in the planet-search by the choice of cluster distance and by
adjusting exposure times for the target.
\end{enumerate}

%---------------------------------------------------------------------------

\section{Main Challenges for Transit Surveys}
\label{mainchallengeTS}

Open cluster planet transit surveys are a subset of planet transit
surveys and therefore have some important challenges in
common. Articulating these challenges is crucial in light of the fact
that over 20 planet transit surveys have been operating for a few
years \citep{h03}, with only six known transiting planets of which
five were discovered by transits.

The most basic goals of any transit survey are to (1) detect planets
and provide their characteristics, and (2) to provide (even in the
case of zero detections) statistics concerning planet frequencies as a
function of the astrophysical properties of the surveyed environment.
Although the most important considerations for designing a successful
transit survey were presented in \citet[M03 hereafter]{msy03} for the
EXPLORE Project, we summarize and provide updates to the three key
issues: number of stars, detection probability, and blending. For OC
surveys, these issues (with the exception of blending) can be
optimized or overcome by a careful survey strategy, particularly by
the selection of the target OC (see \S \ref{targets} through
\ref{observing}).

%---------------------------------------------------------------------------

\subsection{Maximizing Number of Stars with High Photometric Precision}
\label{max_number}

Any survey's goal should be to maximize the number of stars for which
it is possible to detect a transiting planet. We discuss the three
most important aspects below: the astrophysical frequency of
detectable transiting planets, the probability with which existing
planetary transits are observed, and the number of stars for which the
relative photometric precision is sufficiently high.

The frequency of detectable transiting planets is calculated by
considering the astrophysical factors: frequency of CEGPs around the
surveyed stars; likelihood of the geometrical alignment between star
and planet necessary to detect transits; and binary fraction.  We
assume a planet frequency around isolated stars of 0.7\% for planets
with semi-major axis $a \sim 0.05$ \citep{mbf04,nmb04}.
%, and $\sim$ 0.07\% for planets with
%$a \sim 0.025$\footnote{This ``very hot Jupiter'' (i.e., 1- to 2-day
%period planets) frequency is a statistical upper limit, estimated from
%considering the 10 radial velocity planets at $a=0.04$ to 0.05, and
%the one radial velocity planet with $a<0.04$ which may or may not be
%part of the very hot Jupiter class.}.  
Of those CEGP systems,
approximately 10--20\% (probability $\sim R_*/a$) would, by chance,
have a favorable orientation such that a transit is visible from
Earth.  We assume that planets can only be detected around single
stars and conservatively adopt a binary fraction of 50\%; although
there are known planets orbiting binary stars and multiple star
systems \citep{zm02,eum04}, their detection by transits would be
difficult due to a reduced photometric signature in the presence of
the additional star. Combining the above estimates, we arrive at the
at the value of 1 star in 3000 ($a \sim 0.05$ AU)
% or 1 star in 15000 ($a=0.025$ AU) 
having a hot Jupiter planetary transit around a main sequence star.

\begin{figure}
\epsscale{0.5} \plotone{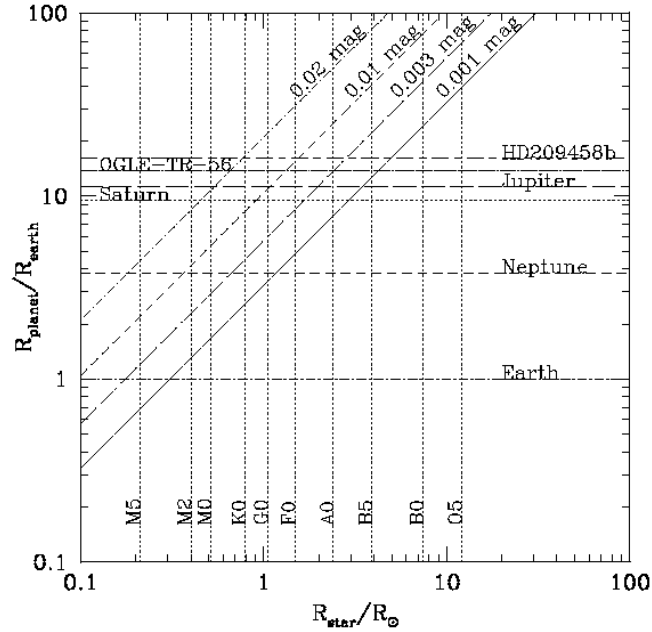}
\caption{Depth of transit signal for transiting planets with different
radii as a function of MK spectral type and corresponding stellar
sizes \citep[from][]{c00} based on geometric arguments only.  The
diagonal lines indicate the amplitude of the transit signal in the
light curve of a given planet--star combination.  For instance, a
Jupiter-sized planet would cause a 0.01 mag dip in the light curve of
a G0 star, but only a 0.003 mag dip in the light curve of an A0
star.}\label{detectability}
\end{figure}

The probability of detecting an existing transiting hot Jupiter
(1/3000) applies only to stars for which it is possible to detect a
planetary transit given the observational setup of the survey.  This
number of ``suitable'' stars is frequently equated to the number of
stars with high enough relative photometric precision to detect the
transiting planet (see Fig. \ref{detectability}).  Planet transit
surveys in general reach photometric precision sufficiently high to
detect Jupiter-sized transiting planets around main sequence stars
(see Fig. \ref{detectability}) for up to 40\% of stars in their
survey, depending on crowdedness and other factors. For example, the
EXPLORE search reached relative photometric precision better than 1\%
on 37,000 stars from $14.5 \leq I \leq 18.2$ out of 350,000 stars down
to $I = 21$ (M03); OGLE-III reached better than 1.5\% relative
photometric precision on 52,000 stars out of a total of 5 million
monitored stars \citep{uzs02}; the Sleuth survey \citep{dck04},
reaches better than 1.5\% relative photometry over the entire dataset
on the brightest 4000 stars out of 10,000 in their 6$\times$6
degree$^{2}$ field, using an automated 10 cm telescope with a 6 degree
square field of view.

The real number of stars suitable for planet transit detection,
however, is not equivalent to the number of stars with 1\% relative
photometry. One may see from Fig. \ref{detectability}, for instance,
that a Jupiter-sized planet would cause a 2\% eclipse around a parent
M0-star. Furthermore, Pepper \& Gaudi (in preparation) find that, if a
planet with given properties around a cluster member star on the main
sequence produces a detectable transit signature, a planet with
identical properties orbiting any other main-sequence cluster member
will produce a detection of approximately the same
signal-to-noise-ratio (SNR) unless the sky flux within a seeing disk
exceeds the flux of the star. Since most transit surveys aim to find
planets of approximately Jupiter-size around stars whose radii are
close to, or less than, a solar radius, the number of stars with 1\%
relative photometric precision can therefore be regarded as a lower
limit to the number of stars suitable for transit detection. For the
rest of this publication, we thus use this number as a proxy for the
number of stars around which we (or other transit surveys) can detect
planets.

%---------------------------------------------------------------------------

\subsection{Probability to Detect an Existing Transit}
\label{pvisi}

The actual observed hot-Jupiter transit frequency will be lower than
1/3000 due to the probability with which an {\it existing} transit
would be observed two or more times during an observing run. This
probability, which we call $P_{vis}$, is equivalent to the window
function of the observations. Although the probability function has
been described in detail before \citep[][and M03]{bs84,g00}, we extend
the discussion to include the recently discovered class of 1-day
period planets, as well as considering different metrics for
probability to detect an existing planet.  In all of our $P_{vis}$
simulations we assume the simplified case of a solar mass, solar
radius star with the planet crossing the star center, thus focusing on
stars we are most interested in. The transit duration is then related
to the planet's period by $t_{\rm duration} = P R_{\odot}/ \pi a$
(typically a few hours for a period of a few days).

In panel a of Figure~\ref{visi} we show the $P_{vis}$ for detecting
existing transiting planets with different orbital periods under the
requirement that two or more full transits must be observed. We
consider different runs (7, 14, 21 nights) of consecutive nights with
10.8 hours of uninterrupted observing each night with 5-min
time-sampling. This is equivalent to approximately 125 observations of
a given cluster per night.  The $P_{vis}$ for 1- to 2-day period
planets is basically complete for the 14- and 21-night runs, while the
$P_{vis}$ is markedly lower for a 14-night run compared to a 21-night
run for planets with periods between 2 and 4 days.

We now turn to $P_{vis}$ for a transit detection strategy where it is
not necessary to detect a transit in its entirety during a single
observing night.  Instead, the strategy requires a transit to be
detected in phased data (from at least two individual transit
events). Such a $P_{vis}$ (which we call $Pvis_{phased}$) is relevant
for transit detections based on period-folding transit-searching
algorithms (for instance with data covering partial nights or a
strategy of alternating targets throughout the course of a
night). With a period-folding algorithm, each individual transit need
not be fully sampled.  In order to quantify $Pvis_{phased}$, we
specify that the phased transit must be sampled by at least $N$
points. Since a typical duty cycle of a transit is on the order of a
few percent, we choose $N$ = 20, 40, or 60 to represent light curves
with a total of a few hundred to a thousand data points (the phased
OGLE planets' light curves typically have a few tens of data points
obtained during transit).  $Pvis_{phased}$ is then calculated to be
likelihood (as a function of period) that at least $N$ in-transit
points are accumulated for observing runs of different lengths and
different observing cadences.

Note that in reality a detection of a planet transit depends on the
number of photons observed during the transiting phase. This number of
photons is contained in the combination of the SNR per individual data
point and the number of data points (during any transit). A
back-of-the-envelope calculation would give a transit SNR for a
$\Delta m \sim$ 2\%-depth transit with $M=20$ data points and a
relative photometry precision of rms $\sim$1\%:

\begin{equation}
SNR \simeq \sqrt{M} \times \frac{\Delta m}{{\rm rms}}
= \sqrt{20} \times \frac{0.02}{0.01} \sim 9.
\end{equation}

For comparison with the two-full-transit $P_{vis}$, we show in Fig.
\ref{visi2} $Pvis_{phased}$ for $N = 20, 40, 60$ by a solid, dotted, and
dashed line, respectively. The four panels represent different
observing strategies: 

\begin{itemize}

\item Panel a shows $Pvis_{phased}$ for 21 nights (10.8h) with
5-minute time-sampling, resulting in a light curve with around 2700
data points. If the data SNR is high enough for 20 data points during
transit to constitute a detection, then $Pvis_{phased}$ is high for
all transit periods between one and five days.  If, in contrast, the
data SNR is lower, and 40 or 60 points per transit are required for
detection, then $Pvis_{phased}$ is low for $P>2$ days.

\item Alternatively, one could imagine a strategy of alternating
between cluster fields (to increase the number of monitored stars), in
which case the observing cadence is reduced. Panel b shows
$Pvis_{phased}$ when observing for 21 nights with a 15-minute cadence
($\sim$ 900 measurements in the light curve). The probability to
detect transits with $N=20$ is very low for $P>2$ days, and it is zero
for $N=40$ or 60.

\item Panel c shows $Pvis_{phased}$ for 40 nights of continuous
observing (10.8h) with an observing cadence of 5 minutes ($\sim$ 5200
data points). For $N=20$ and $N=40$, $Pvis_{phased}$ is close to
complete for all shown periods.

\item Panel d shows $Pvis_{phased}$ for 40 nights of observing with a
15-minute cadence (again simulating a strategy of alternating between
cluster fields; $\sim$ 1700 data points). $Pvis_{phased}$ is very low
for $N>20$, indicating that, in order to be able to observe with a
15-cadence, many more than 40 nights are needed if more than 20 data
points are required for a transit detection.

\end{itemize}

We conclude that the ability to particularly detect longer-period ($P
>$ 2 days) planets depends on observing strategy. For the rest of this
paper we adopt the $P_{vis}$ criterion of seeing two full transits
which is a good strategy for a limited number ($\sim 20$) of observing
nights.

%---------------------------------------------------------------------------

\subsection{Blending and False Positives}\label{blending}

%There are many false positive planet transit candidates found in
%planet transit searches, including grazing binary stars and eclipses
%of M dwarfs orbiting larger stars, as well as blending.

Blending in a planet transit light curve due to the presence of an
additional star is a serious challenge inherent in planet transit
surveys, one that has only recently been gaining recognition. If the
light of multiple stars are interpreted as being due to one individual
star, then the relative depth of any eclipse will be decreased. This
``light pollution'' may either cause (1) an eclipsing binary systems
to mimic a more shallow transiting planet signal, or (2) a true {\it
planet's} transit signal's depth to be decreased to a fraction of its
already very small amplitude, rendering it harder or even impossible
to detect.

Blending can be caused either by optical projection in crowded fields,
or by physically associated stellar systems. The crowdedness can be
considered in the choice of target and observational setup. Blending
in spatially unresolved, physically associated systems generally
consists of a wide binary of which one component hosts an additional
close-by stellar companion or a transiting planet.  The component
without the close-by stellar or planetary companion would produce the
polluting light. Although we have accounted for binary stars in our
probability estimate (\S \ref{max_number}), the contribution due to
this kind of ``false positive'' may be larger due to the unknown
wide-binary component distances.

Recently the effect of blending on the probability of detecting
planets has been addressed by several authors \citep[e.g.,
M03;][]{b03,dck04,kts03} in the context of causing false
positives. Several solutions have been proposed to avoid false
positive transit candidates that are actually blended star light
curves. \citet{sm03} show that one may eliminate some false positives
due to blending with photometric data alone if the light curve is of
sufficient relative photometric precision and the observing cadence is
high enough to clearly resolve the individual temporal components of
the transit. Using spectroscopic data, other solutions include a
careful modeling of the additional star properties to detect a second
cross correlation peak caused by a physically associated star
\citep[M03;][]{kts03,kcl03,tks04a}. Finally, estimates for blending
(associated either with chance alignment of foreground or background
stars, or physical triplets) effects on the probability of detecting
existing transits can be quantified for individual surveys, as done by
\citet{b03} for shallow wide-field transit surveys.

%While attention has been given to identifying false positives, more
%emphasis is needed on the significance of blending on lowering the
%number of planets detectable by transits. In particular, 

%lower Open cluster surveys will have fewer
%contaminating giant stars than the shallow transit surveys and are
%less likely to miss existing transits as a result of blending due to
%the smaller confusion radius (typically more on the order of 1''). Our
%probability estimates are more optimistic (\S \ref{max_number}) since
%we do not include easily identified false positives in our estimate.

%---------------------------------------------------------------------------

\section{Main Challenges for Open Cluster Transit Surveys}
\label{mainchallengeOC}

\noindent The difficulties and challenges involved in searching for
planetary transits specifically in OCs are: the fixed and somewhat low
number of stars in an open cluster, determining OC cluster membership
in the presence of significant contamination, and differential
reddening along the cluster field and along the line of sight.  We
outline these aspects individually below.

\begin{enumerate}

\item
{\bf The Number of Monitored Stars} The number of monitored stars is
typically lower than in rich Galactic fields (in part due to the
smaller field size of the used detectors), reducing the statistical
chance of detecting planets. Open clusters can have up to $\sim$
10,000 member stars \citep{f95}, depending on the magnitude range
taken into consideration. Only a subset of these stars, perhaps
10--20\%, however, will be observed with sufficient relative
photometric precision to detect transits (cf. \S
\ref{max_number}). The number of these stars in rich OCs is comparable
to the number in wide-field, shallow transit surveys, e.g., Sleuth:
$\sim$4000 stars in 6$\times$6 square degrees of $9 < R < 16$
\citep{dck04}; WASPO: $<$ 3000 stars in 9$\times$9 square degrees of
broadband magnitude between 8 to 14 \citep{kch04}. The richest deep
galactic fields surveyed have many more stars with high relative
photometric precision (\S \ref{max_number}).

\item
{\bf Cluster Contamination} Determining cluster membership of stars in
the OC fields without spectroscopic data or proper motion information
is difficult due to significant contamination by Galactic field stars,
since the clusters are typically concentrated toward the Galactic
disk.  For example, \citet{shl03} estimate the contamination of
Galactic field stars in their study of NGC 6819 to be around 94\%.  A
study by \citet{nsp02} calculated the average contamination in the
fields of 38 rich OCs to be 35\% in the inner regions and 80\% in the
``coronae'' of the clusters. Furthermore, if the target is located
such that the line of sight includes a long path through the Galaxy
(e.g., low Galactic latitude and longitude towards the Galactic
bulge), background giants may start polluting the sample of stars with
apparent magnitudes monitored with high relative photometric precision
\citep[see for example the discussion in][]{shl03}. Getting a handle
on the issue of contamination is vital for OC surveys since any
statistical statements about the result will need to be based on
estimates of surveyed cluster members.

\item {\bf Differential Reddening} Differential reddening across the
cluster field and along the line of sight can make isochrone fitting
(and subsequent determination of age, distance, and metallicity)
difficult. Ranges of $\Delta E_{B-V} \sim 0.2$ or higher across
fields-of-view of 10--20 arcmin on the side are not uncommon
\citep[see for example the studies
by][]{mc96,rcb97,rb02,cm04,vbc04,pcs04}. For an $R_{V}=3.1$
reddening-law, a differential $\Delta E_{B-V} \sim 0.2$ corresponds to
a differential $\Delta V \sim 0.6$ and $\Delta I \sim 0.3$
\citep{ccm89}. The calculated effective temperature of a
solar-metallicity main-sequence star with $V-I \sim 0.8$ would vary by
about 500 K for a differential reddening effect of $\Delta E_{B-V}
\sim 0.2$ \citep{hbs00}.  It should, however, also be noted that some
OCs do not seem to suffer from differential reddening, such as NGC
1245 as examined in \citet{bgd04} and NGC 2660 in our preliminary
analysis of its CMD.

%SS: I cut out the following since we are focusing on differential
%reddening:
%Furthermore, literature data on OCs, which can be used
%to constrain one or several free parameters in the isochrone fitting
%process, are sparse and/or different publications may give conflicting
%results as we describe in more detail in Sections \ref{targets} and
%\ref{targets_exp}.

\end{enumerate}

%---------------------------------------------------------------------------

\section{Open Cluster Selection}\label{targets}

Open cluster target selection can help overcome or reduce some of the
main challenges of OC planet transit surveys described in
\S\ref{mainchallengeTS} and \S\ref{mainchallengeOC}. More
specifically, careful cluster selection can help maximize the number
of stars, maximize the probability of detecting existing transits, and
reduce line-of-sight and differential reddening.  Most importantly,
cluster selection allows for targeting specific spectral type for a
given telescope and observing cadence. 

The biggest challenge in the selection of target clusters is the
paucity of data about many OCs. The physical parameters of the
cluster, such as distance, foreground reddening, age, and metallicity
are frequently either not determined, or there exist large
uncertainties in the published values. For example, out of
approximately 1100 associations of stars designated as OCs, many only
have identified coordinates, approximately half have an established
distance, and about 30\% have an assigned metallicity \citep[WEBDA
database][]{m96}. Additional difficulties arise when independent
studies arrive at different values for any of the parameters. For
instance, the metallicity of NGC 2660 was determined to be as low as
-1.05, and as high as +0.2 \citep[see discussion in the introduction
of][]{sbt99}. The very important criterion of richness tends to be
even less explored than the other physical parameters, probably due to
the significant contamination from field stars that these OCs tend to
suffer.

In spite of the lack of OC data, a very useful place to start is the
WEBDA\footnote{\url{http://obswww.unige.ch/webda/}} database
\citep{m96}. From the long list of potential OC monitoring targets,
one can then start eliminating cluster candidates by applying the
criteria we describe below.

%---------------------------------------------------------------------------

\subsection{Cluster Richness and Observability}\label{richness}

Apart of its observability for a given observing run\footnote{To
optimize this observability for our potential cluster targets, one may
use \texttt{SKYCALC}, written by J. Thorstensen and available at
\url{ftp://iraf.noao.edu/iraf/contrib/skycal.tar.Z}}, the most
important selection criterion for a cluster is its richness, simply to
increase the statistical chance of detecting planets.  The richness of
the cluster field can be estimated by looking at sky-survey
plots\footnote{Available, for instance, at
\url{http://cadcwww.hia.nrc.ca/cadcbin/getdss}} of the appropriate
region.  Estimating the richness of the cluster itself is a much more
difficult process since field star contamination is usually
significant due to the typically low Galactic latitude of the Pop. I
OCs \citep[see below and][]{bhb03,shl03,bgd04}. One may use published
cluster richness classifications, such as in \citet{c00}, taken from
\citet{ja82}. The WEBDA database also contains information for
clusters in the 1987 Lynga catalog \citep[online data published
in][]{l95}. The data published there, however, only present lower
estimates for richness classes. The depths of the studies from which
the richness classes were derived may differ significantly from one
study to the next. Thus, these classes are rough estimates only, and
the best way to judge the richness of a cluster field is to rely on
one's own test data obtained with the same setup as the one used for
the monitoring study (see \S \ref{contamination}).

\subsection{Cluster Distance}\label{distance_reddening}

The distance to the target cluster is an important criterion for
cluster selection for four reasons: (1) to ensure the cluster is
sufficiently distant to fit into the field of view, (2) to allow RV
follow-up of potential candidates, (3) to target the desired range of
spectral types for given observing conditions, and (4) to minimize
reddening.

Since all stars in an OC are at approximately the same distance, one
may, with appropriate adjustment of the exposure time for given
telescope parameters, cluster distance, and foreground reddening,
target certain spectral types of stars for high-precision photometry.
We are interested in G0 or later spectral type/smaller stars, since
early-type stars have larger radii which would make transit detections
more challenging. The transit depth in the light curve is, for small
$\Delta m$, simply given by
\begin{equation}
\Delta m \simeq \frac{\Delta F}{F_0} =
\left(\frac{R_{planet}}{R_{star}}\right)^2,
\end{equation}
\noindent where $\Delta m$ and $\Delta F$ are the changes in magnitude
and flux, respectively, and $F_0$ is the out-of-transit flux of the
parent star (see Fig. \ref{detectability}). In addition to featuring
larger radii, early spectral types are fast rotators which exhibit
broad spectral lines, making mass-determination of planetary
companions more difficult.

Foreground reddening, increasing with cluster distance, usually
represents a proxy for the amount of differential reddening across the
field of view as well as along the line of sight \citep{sfd98}.
Differential reddening will cause the main sequence of the cluster to
appear broadened \citep[e.g.,][]{bm01}. This, in turn, will cause
large errors in the determination of cluster parameters such as age,
metallicity, etc, by means of isochrone fitting. Furthermore, a broad
main sequence will make any attempts to estimate contamination
\citep[based on isochrone fitting; see for example][]{msf98,hss02}
more challenging.

%---------------------------------------------------------------------------

\subsection{Cluster Age}

The consideration of cluster age in the OC selection is not as crucial
as richness, observability, and distance. However, choosing both
younger and older OCs may impose different challenges with respect to
the transit-finding process.

Stellar surface activity, which would introduce noise into the light
curve of a given star, decreases with age.  As they age, stars lose
angular momentum and thus magnetic activity on their surfaces
\citep[see for example][and references therein]{d98,w04}.  The
photometric variability for a sample of Hyades OC stars was found to
be on the order 0.5--1\% \citep{psc04} with periods in the 8--10 day
range \citep[the Hyades cluster has an age of $\sim$650 Myr;
see][]{pbl98,lfl01}. While these photometric variations do not
necessarily represent a source of contamination in the sense of
creating false positives, they nevertheless will introduce noise into
the stellar light curves and thus render existing transits more
difficult to detect.

Stellar surface activity is potentially an issue not just for host
star photometric variability, but more so for radial velocity
follow-up. \citet{psc04} found that radial velocity rms due to
rotational modulation of stellar surface features can be as high as 50
m/s and is on average 16 m/s for the same sample of Hyades stars.
Further such correlation between radial velocity rms and photometric
variability was found by \citet{qhs01} who observed a $v_r$ amplitude
of $\sim$ 180 m/s for HD 166435 (age 200 Myr), a star without a
planet. The associated photometric variability with a period of around
3.8 days is of order 5\%.

This variability issue favors older OCs as targets, particularly since
most of the decrease in surface activity occurs between stellar ages
between 0.6 Gyrs and 1.5 Gyrs \citep{pp04}. A 16 m/s rms may not be a
problem for deep OC surveys; for short-period Jupiter-mass planets
relatively large radial velocity signatures are expected and the faint
stellar magnitudes limit radial velocity precision to $\sim$ 50--100
m/s \citep{ktj03, kts04, bps04} using currently available telescopes
and instrumentation.

Older star clusters offer an additional advantage: in general older
OCs are richer and more concentrated, and therefore offer a larger
number of member stars to be surveyed \citep{f95}. On the other hand,
some old OCs appear to be dynamically relaxed and mass-segregated
\citep[such as NGC 1245;][]{bgd04}, and in the case of NGC 3680, for
instance, evidence seems to point toward some resulting evaporation of
low-mass stars over time \citep{naa97}.  Since low-mass stars are the
primary monitoring targets, some dynamically evolved OCs may actually
be less favorable for observing campaigns.

%---------------------------------------------------------------------------

\subsection{Other Criteria}\label{further}

% this was too wordy. i made it more concise
%Further criteria which may influence the choice of observing target
%are the ones listed below. We note, however, that these may be more
%difficult to take into account since one may be limited to a smaller
%number of observing targets after allowing for the selection criteria
%described above.
Given a sufficiently high remaining number of suitable open clusters
after consideration of the previous four selection criteria, cluster
metallicity and galactic location are additional relevant selection
criteria.

\begin{itemize}

\item Range of metallicities. In order to be able to make quantitative
statements about planet frequency as a function of metallicity of the
parent star (\S\ref{transits_ocs}), one needs to have a sample of
clusters with varying metallicities. Surveys based on the
radial-velocity method indicate that solar-neighborhood stars with
higher metallicities are more likely to harbor planets than metal-poor
ones \citep{fv03}. It may therefore be advantageous to favor
higher-metallicity clusters for monitoring studies to find planets.

\item The target's Galactic coordinates. On average, the closer the OC
is to the Galactic disk, the higher the contamination due to Galactic
field stars (see \S \ref{mainchallengeTS}). Moreover, if the target is
located close to, or even in front of, the Galactic bulge,
contamination may be severe. It should be pointed out that background
giants and subgiants will truly pollute the stellar sample since their
radii are too large to detect planets. Transit detections around
main-sequence field stars at distances of less than, or roughly equal
to, the OC distance are still possible and would be as scientifically
valuable as a detection of a planetary transit as part of a dedicated
field survey.

\end{itemize}

%---------------------------------------------------------------------------

\section{EXPLORE/OC Target-Selection Strategy}
\label{targets_exp}

EXPLORE/OC is a transit survey in open clusters operating with the LCO
1m Swope Telescope with a field of view of 24 arcmin $\times$ 15
arcmin and a scale of 0.435 arcsec/pixel. We have observed 5 open
clusters to date: NGC 2660 observed for $\sim$ 15 nights in Feb 2003
\citep{blm04}; NGC 6208 observed for $\sim$ 21 nights in May-June 2003
\citep{lbm04}; IC 2714 observed for $\sim$ 21 nights in March/April
2004; NGC 5316 observed for $\sim$ 19 nights in April 2004; and NGC
6253 observed for $\sim$ 18 nights in June 2004\footnote{We define
here the number of nights as the number of at least partially useful
nights during the monitoring campaigns.}.

Our $I$-band, high-cadence ($\sim$ 7 min, including 2 min readout
time) photometric monitoring enables us to typically attain 1\%
precision in our relative photometry for around 3000--5000 stars per
cluster target field in the range $14.5 < I < 17$ (see
Fig. \ref{rms}). This number corresponds to a lower limit to the
number of stars around which we can detect planetary transits (see \S
\ref{max_number}).  In the context of outlining our survey strategies,
we present some of our preliminary results of the studies of the open
clusters NGC 2660 and NGC 6208. In this Section we explain our
approach to target selection, specifically designed to maximize the
number of target stars of appropriate spectral type.

%---------------------------------------------------------------------------

\subsection{Overall Potential Targets}

Our potential OC targets listed in Table \ref{potential} were chosen
with the basic goal that we observe as many cluster member stars as
possible at a sufficiently high photometric precision and high-cadence
of observations to detect CEGPs around them. Richness classes are
given whenever they were available. We note that we used these
published richness classes only as a guideline (i.e., we gave extra
considerations to OCs classified as rich, but did not necessarily
discard any OCs classified as poor) and relied more on visual
inspection and photometric analysis of sky survey images of the
cluster regions.

Targets in Table \ref{potential} were further selected based on the
published estimates for distance and foreground
reddening\footnote{Note that our first target, NGC 2660, selected on
the basis of its estimated richness and observability alone, turned
out to have a relatively large distance and high foreground
reddening.}. To select a cluster with a suitable distance we consider
the preferred range of spectral types (G to M), our adopted relatively
short exposure times (see \S \ref{cadence}), and the size of the LCO
Swope telescope.

As an example of how distance, exposure time, target spectral type,
and reddening are related we use our OC NGC 6208.  In Fig. \ref{rms},
we show our photometric precision as a function of $I$ magnitude of
our NGC 6208 data (night 15), obtained during May and June 2003 at the
LCO 1m Swope Telescope. We conservatively estimate that, with our
exposure time of 300s per frame, we attain 1\% precision for a range
of about 2.5 magnitudes ($14.5 < I < 17$). From the WEBDA database
(see also Table \ref{potential}) we find that the distance to NGC 6208
is 939 pc, and the foreground reddening is $E_{B-V} = 0.210$.  Using
the relation $A_{I} = 1.94 E_{B-V}$ from \citet{sfd98}, we find that
$I = 17$ for NGC 6208 cluster members corresponds to $M_{I} = 6.73$
which we call $M_{I_{limit}}$ in Table \ref{potential}. Using table
15.7 in \citet{c00}, this corresponds to an MK spectral type of M0 or
M1.  The bright limit, above which saturation will start to set in,
would be at $M_{I} \sim 4.2$ which would correspond to an MK spectral
type of approximately G5 \citep{c00}.  Once the range of spectral
types for the monitored cluster members is determined, we can estimate
the range of planetary radii which would be detectable (see
Fig. \ref{detectability}).

%---------------------------------------------------------------------------

\subsection{Potential Targets for a Given Observing Run}

For a given slot of observing time, we use Table \ref{potential} as
the source from which we pre-select two or three potential observing
targets for the run. The final target selection is then performed
based on our own data, taken either during a previous observing run or
at the very beginning of the observing run itself (see below). The
main criterion at the pre-selection stage is the observability of the
potential targets to maximize the time during which we can observe the
OC.

With all of the other constraints (distance, reddening, richness) on
cluster selection, finding a cluster that is observable all night long
becomes challenging when observing runs are long. The main criterion
for a successful transit search is maximizing the time during which we
can observe the respective target OC (to increase $P_{vis}$), making
clusters of numerically high southern declination preferable targets.

%---------------------------------------------------------------------------

\subsection{Final Target Selection}\label{final_selection}

%I cut the following since we don't need a summary here.
%As we mentioned above, our approach is to pre-select targets based on
%the various strategies described above, which first resulted in the
%list in Table \ref{potential}, and then select a subset thereof, based
%on the allocated observing time and OC coordinates. We then obtain our
%own test data of the targets within this subset, either during an
%observing run prior to the actual monitoring study of the eventual
%target, or during the first night of the monitoring run itself. 
Our final target selection is based on the evaluation of $VI$ test
data (see Fig. \ref{final_target_selection}) for the group of
pre-selected clusters, involving the following steps:

\begin{enumerate}
\item We create (at least roughly) calibrated $VI$ CMDs of the
potential target clusters based on our own test data. These data were
obtained either during the beginning of the same observing run or
during prior runs during photometric conditions and reasonable seeing,
and have the same exposure time as for the eventual monitoring. Figure
\ref{final_target_selection} shows these CMDs for the OCs NGC 6253
(left panel) and NGC 6134 (right panel), both of which were targets
for our June 2004 run.
\item Within this CMD, we count the number of stars for which we
expect to obtain photometry down to 1\% or better, which, according to
Fig. \ref{rms}, will include most stars with $14.5 < I < 17$. Note
that we pre-select our targets based on their distance, so that stars
within this range of apparent magnitude will be of spectral type G or
later.
\item As a final step, we perform cuts in $V-I$ color to eliminate the
redder sequence of background evolved disk stars if it is present in
the CMD. Stellar radii of evolved stars are significantly larger than
their main-sequence counterparts, and thus detecting planets around
evolved stars is virtually impossible due to the reduced photometric
signal depth of a transiting planet. We show how we eliminate the
evolved sequence from consideration in the left panel of Fig.
\ref{final_target_selection}.
\item The result of this count approximately corresponds to the number
of small main-sequence stars we can monitor at the 1\% photometry
level and serves as the figure of merit in the cluster selection
decision-making process. Since the box in the CMD of NGC 6253 contains
more stars (3400) than the one for NGC 6134 (2850), NGC 6253 was
chosen as our observing target for June 2004. The last column of Table
\ref{potential} shows the estimates of the numbers of 1\%-rms stars
for our potential target clusters which have test data available.
\end{enumerate}

%---------------------------------------------------------------------------

\section{EXPLORE/OC Observing Strategy}\label{observing}

The EXPLORE/OC observing strategy is designed to maximize $P_{vis}$,
minimize false positives, and to constrain field contamination---the
issues described in \S \ref{mainchallengeTS} and
\ref{mainchallengeOC}.  We review aspects of observing strategy that
are most important for our project. Some of these are covered in M03
but are included here for completeness. We focus in particular on
considerations necessitated by observing OCs instead of Galactic
fields.

%---------------------------------------------------------------------------

\subsection{Choice of Filter}\label{filters}

Our photometric monitoring is done in the $I$-band. The shape of a
transit in the photometric light curve is dependent on the filter due
to the color dependence of limb darkening whose effects are smaller in
$I$ than in the bluer bands. (see \S 2 and figure 2 in M03).  The
transit depth is near constant in $I$ when the planet is fully
superimposed on the parent star. Because of this ``flat-bottomed''
light curve in $I$, the shape of the transit makes it easier to
distinguish planet transits from the signal caused by grazing binaries
(basically a `pointy' or `round' eclipse instead of a flat-bottomed
one) than at bluer bands where limb darkening is
stronger. Fig. \ref{lightcurves} shows a light curve with a
flat-bottomed eclipse, illustrating that flat bottoms do indeed occur
at I-band.

Additional advantages of observing in the $I$-band are (a) increased
sensitivity to redder, intrinsically smaller stars which offer greater
chances of detecting orbiting CEGPs, and (b) suffering less extinction
due to dust than in the bluer bands.

Disadvantages may include (a) lower CCD quantum efficiency in the
$I$-band compared to, e.g., the $R$-band, and (b) the occurrence of
fringing due to multiple reflections and subsequent interference
internal to the CCD substrate or between the supporting substrate and
the silicon. Fringing is usually more visible in $I$ than in $BVR$,
due to the abundant night sky emission lines in the $I$ wavelength
range.  We note that we do not encounter any fringing at all with our
setup at the Swope Telescope at LCO.

We also do not change filters during OC monitoring since such a
strategy would effectively reduce our observing cadence (\S
\ref{cadence}).

%---------------------------------------------------------------------------

\subsection{Single-Cluster/High-Cadence Observing}\label{cadence}

In order to maximize the chance of detecting any existing planetary
transits, we do not alternate OC targets (even though we would
increase the number of monitored stars that way) but instead observe
the same cluster for as many hours as possible during the night
(Figures \ref{visi2} and \ref{comparison}). The main reason for this
strategy is to conduct high-cadence observing.

The main goal of this approach is to distinguish a true transit light
curve from false positives such as grazing eclipsing binary stars, an
M-star eclipsing a larger star, or stellar blends \citep{sm03,cbd04}.
Because the total duration of a short-period planet transit is
typically a few hours, with ingress and egress as little as 20
minutes, high-cadence observing is essential for well-resolved light
curves for a limited-duration observing run where only two or three
transits are expected.  A well-resolved light curve with good
photometric precision can be used to derive astrophysical parameters
of the planet-star system from the light curve alone
\citep[e.g.,][]{sm03} which is useful in both ruling out subtle false
positives such as blended eclipsing binaries, and in obtaining an
estimate of planet radius.  In particular, the density of the parent
star is of interest for distinguishing between a planetary transit in
front of a main-sequence star, and the case of a late-type dwarf
orbiting a giant star. The star's density, however, (1) can only be
calculated from photometry data alone when assuming a stellar
mass-radius relation, and (2) is sensitively dependent on the full
duration of the transit (including ingress and egress), and the
duration of totality only.

The flatness of a light curve during the out-of-eclipse stages of a
system offers another means of separating planetary transits from
stellar eclipses, as illustrated in \citet{sp03} and
\citet{d03}. Short-period binary stars will have gravitationally
distorted, non-spherical shapes which will result in a constant
sinusoidal brightness variation of the light curve with a maximum at
quadrature.

%We note that high-cadence observing is not required to discriminate
%against false positives as long as a sufficient number of data points
%at all phase angles is obtained during a long observing run. In other
%words, if enough points during planet transit phased over all data are
%obtained, the shape of the light curve can be mapped out. In
%Figure~\ref{visi2} we show the probability (averaged over all phases)
%that at least 10, 20, or 30 points during transit can be obtained for
%periods of 1 to 5 days for different observing run lengths and
%observing cadence.
%
%Exposure time, target-star spectral types, a cluster's distance, and
%foreground reddening are interrelated. For 300s exposure time at the
%LCO 1m in photometric conditions, we typically obtain the necessary
%precision in the relative photometry for stars with $14.5 < I < 17$
%(Fig. \ref{rms}). Since we are interested in stars with spectral type
%G or later, we preferentially target clusters with a distance of around 1
%kpc, given a ``generic'' $E_{B-V} \sim 0.2$.

%---------------------------------------------------------------------------

%\subsection{Single Cluster Observing}\label{pvis}

\begin{figure}
\epsscale{0.5} \plotone{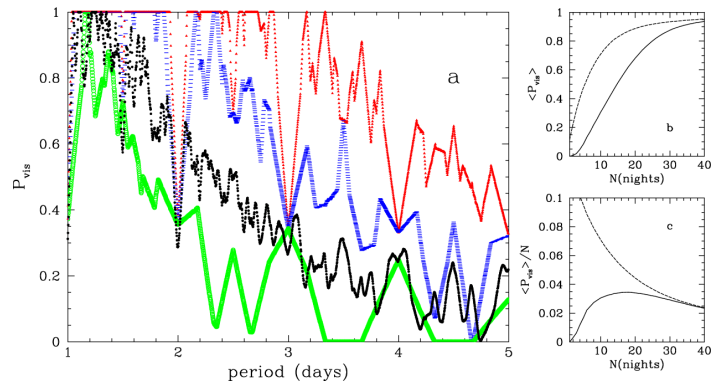}
\caption{\label{visi} Probability $P_{vis}$ of detecting existing
transiting planets with different orbital periods. $P_{vis}$ is
calculated with the requirement that two full transits must be
observed.  {\bf Panel a}: $P_{vis}$ of detecting 2 transits of an
existing transiting planet with a period between 1 and 5 days after 21
(top curve), 14 (second curve from the top) and 7 (bottom curve)
consecutive, uninterrupted nights of observing (10.8 hours per
night). The difficulty of detecting some phases is shown by the dips
in the curves (e.g., orbital periods of an integer number of days may
always feature their transits during the day and are therefore
statistically harder to detect). All phases are averaged over for each
period. The second curve from the bottom shows the real $P_{vis}$ for
our monitoring study of NGC 2660 (19 nights of 7-8 hours per night,
with interruptions due to weather and telescope scheduling; see
Fig. \ref{lightcurves}).  {\bf Panel b:} The mean $P_{vis}$ (averaged
over 1 day $< P <$ 5 days) as a function of number of consecutive
nights in an observing run. The solid line is for the requirement to
detect two transits and the dashed line for one transit. This figure
indicates how much the likelihood of finding existing transits grows
with an increasing number of nights of observing.  {\bf Panel c:} Run
efficiency (defined as $<P_{vis}>$ divided by the number of observing
nights) as a function of run length. For the two-transit requirement
(solid line) and, an observing run of 18 nights is most efficient. For
the single transit requirement, the efficiency decreases monotonically
with the number of nights since additional nights have progressively
lower probabilities of detecting "new" transits.}
\end{figure}

\begin{figure}
\epsfig{file=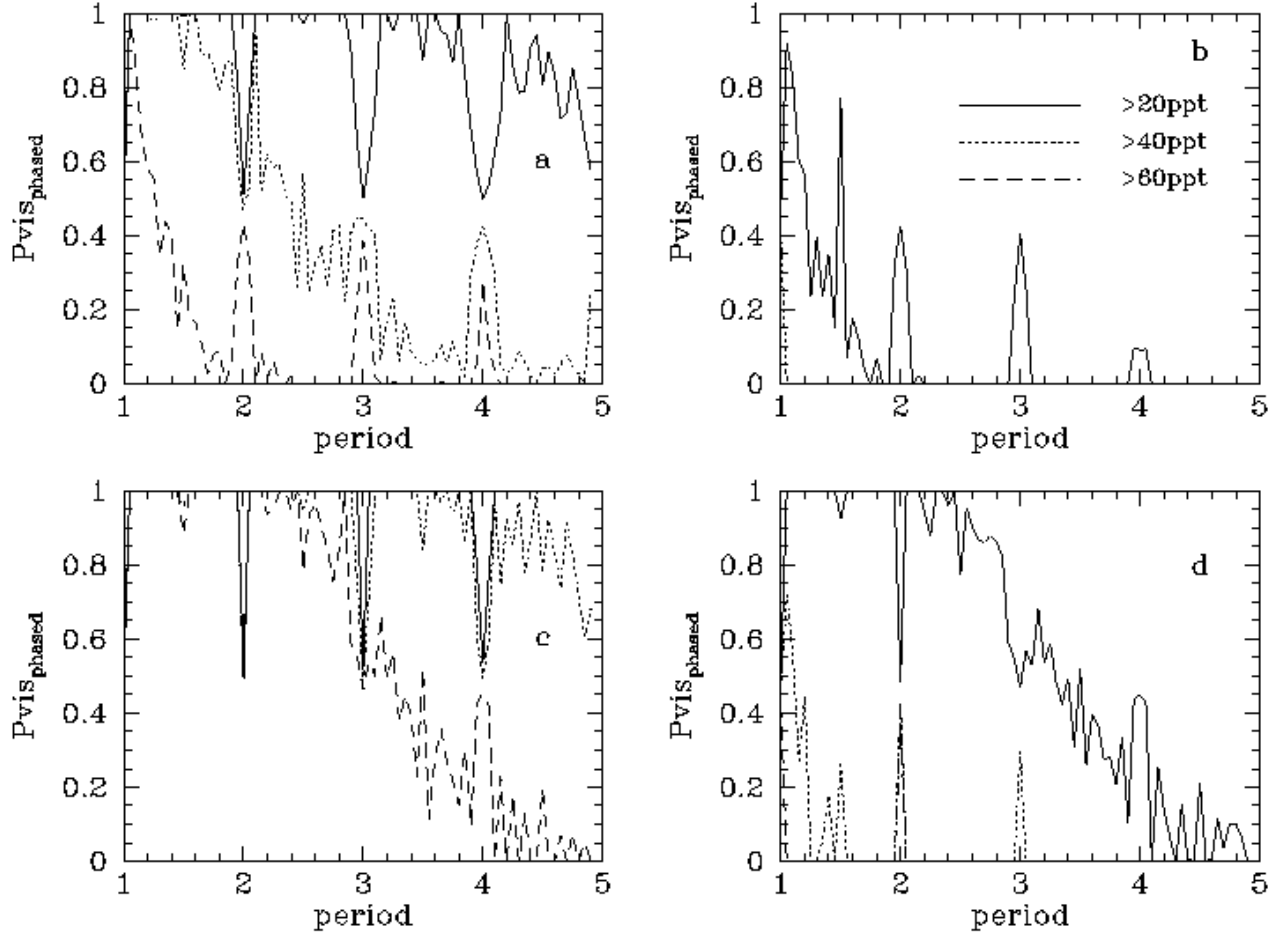, angle=-90, width=\linewidth}
\caption{$Pvis_{phased}$ as a function of period (in days) for
detecting planetary transits in phased data.  $Pvis_{phased}$ is
calculated to be likelihood that at least $N=20$ (solid line), 40
(dotted line), or 60 (dashed line) in-transit points are are
accumulated for observing runs of different lengths and different
observing cadences.  The number of points per transit required for a
detection is dependent on both SNR and exposure time. Panels a and b
compare $Pvis_{phased}$ for a 21-night (10.8h) observing run with a
cadence of 5 minutes (panel a) and 15 minutes (panel b).  Panels c
(5-minute cadence) and d (15-minute cadence) illustrate the same
for a observing run of 40 nights.  See text (\S \ref{pvisi}) for
discussion.}\label{visi2}
\end{figure}

\begin{figure}
\epsfig{file=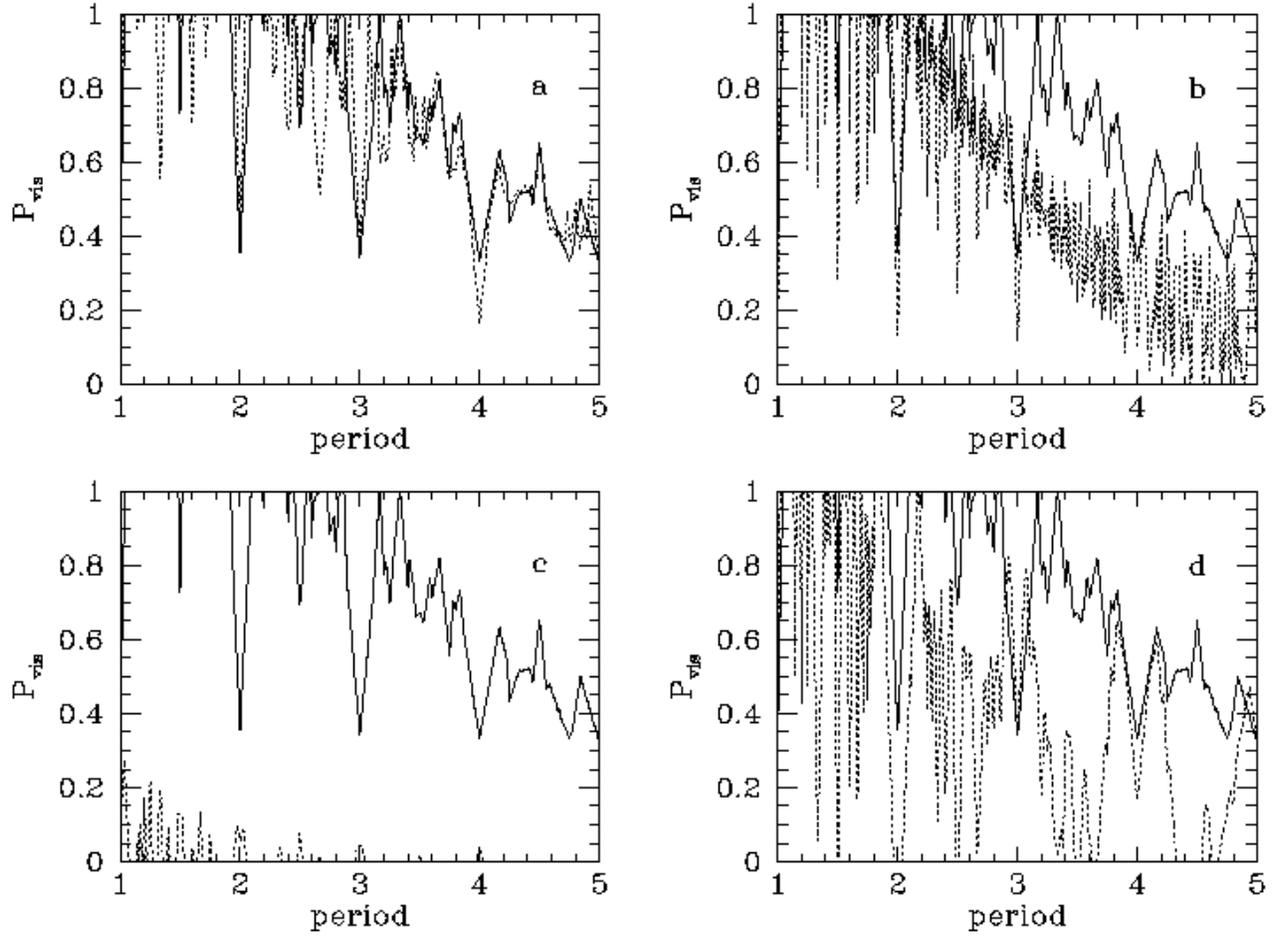, angle=-90, width=\linewidth}
\caption{Dependence of $P_{vis}$ upon period (days) when using
different observing strategies.  The solid line in all four panels
corresponds to $P_{vis}$ (two full transits) of a 20-night (10.8h)
uninterrupted observing run. In {\bf Panel a}, the dotted line
corresponds to $P_{vis}$ of a 40-nights observing run during which we
only observe the first two out of every four nights. $<P_{vis}>$
(periods between 1 and 5 days) of the 20-night observing run is
0.681. $<P_{vis}>$ for the 2-nights-on, 2-nights-off strategy over 40
nights is 0.666.  The dotted line in {\bf Panel b} shows $<P_{vis}>$
when observing the first half of every night for 40 nights in a row:
$<P_{vis}> \sim$ 0.437. When observing a third of every night for 60
nights, as shown by the dotted line in {\bf Panel c}, $<P_{vis}>$ goes
down to 0.007. Finally, {\bf Panel d} illustrates the aliasing effect
of only observing 2 out of 4 nights. The dotted line represents the
probability of two observed transits being consecutive, as a function
of period.  Averaged over all periods, only about half of all detected
pairs of transits would be consecutive when following the 2-nights-on,
2-nights-off strategy.  Note that these numbers are slight
overestimates (few percent) because they do not account for the drift
of sidereal time that would affect a specific target's
observability. For details, see text (\S
\ref{observing}).\label{comparison}}
\end{figure}

\begin{figure}
\epsscale{0.5} \plotone{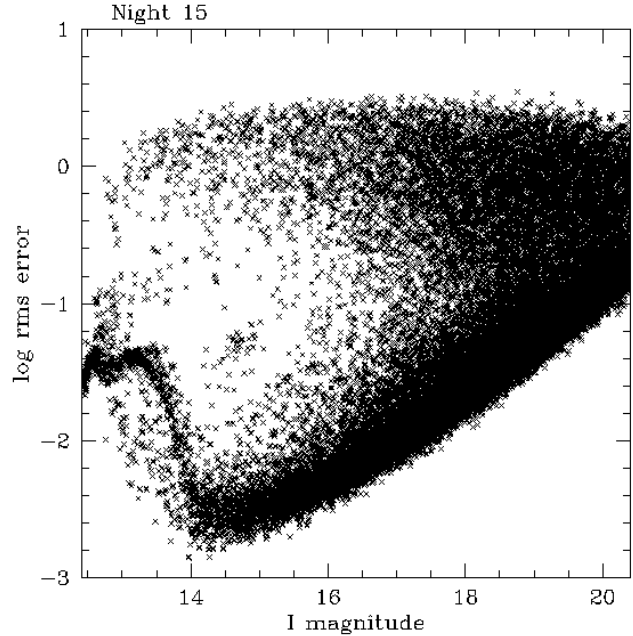}
\caption{\label{rms} Photometric precision of night 15 of our
monitoring run of NGC 6208. In this diagram, slightly more than 5000
stars have photometry of precision 1\% or better. This rms is measured
as the scatter around the mean magnitude of the star under
investigation. The 1\%-photometry stars cover a magnitude range of
slightly more than 2.5 mags. The ``Z''-shaped feature for stars
brighter than $I \sim 14$ is due to the onset of saturation for some
of the stars in some of the images in the time series. The clustering
of stars around log rms $\sim$ 0 is caused by crowding effects when
for some of the images, faint stars are blended together with nearby
bright stars.
%this didn't seem relevant to this caption:
%By adjusting the exposure time, 
%for a given telescope and fixed cluster distance and reddening, one can
%target OC member stars of a range of certain spectral types
%and stellar radius to maximize the likelihood of detecting a transit.
}

\end{figure}

\begin{figure}
\epsscale{0.9}
\plottwo{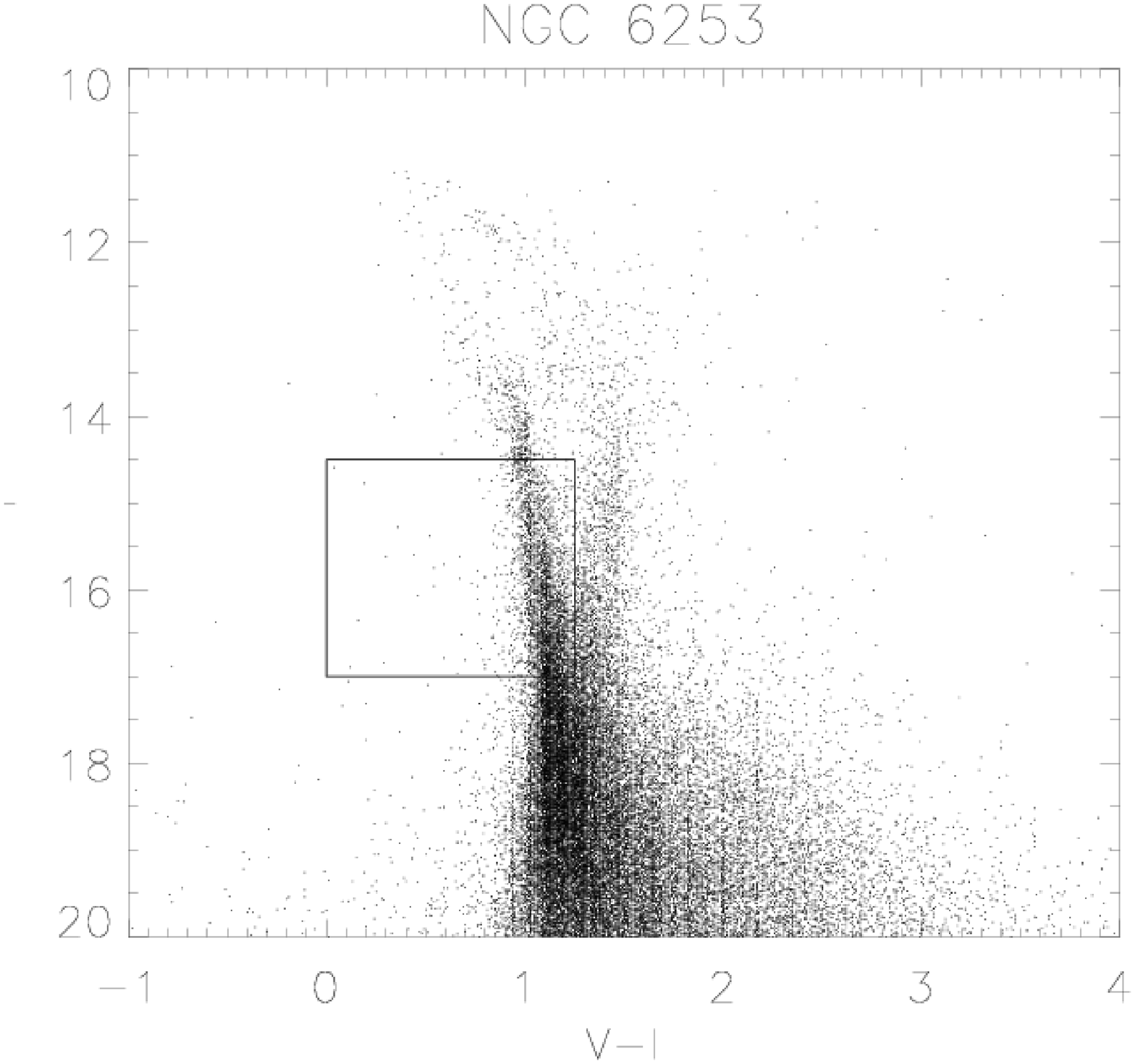}{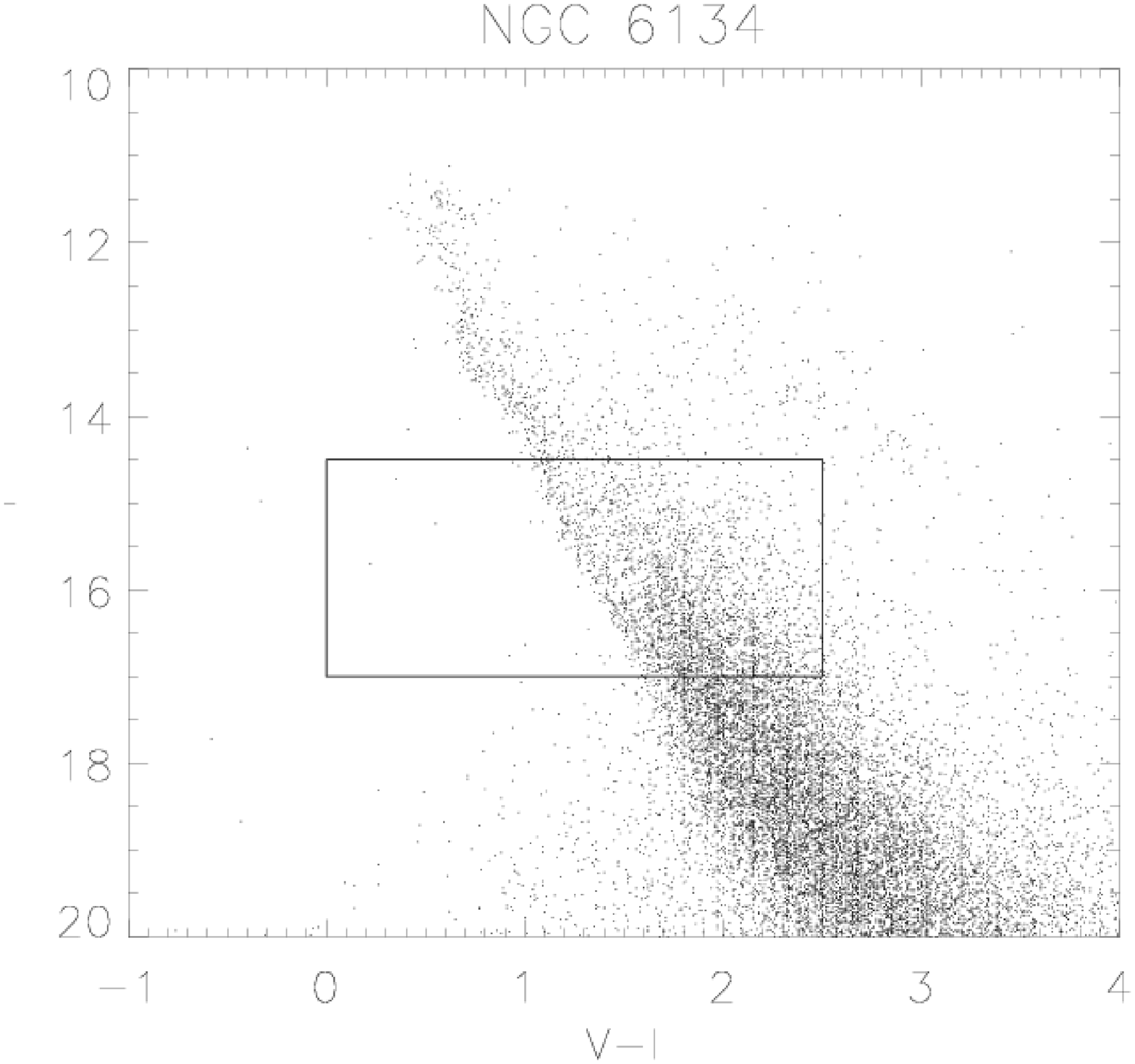}
\caption{This figure illustrates our approach concerning final target
selection, based on our own test data. To choose between the two
potential target clusters of different distances and foreground
reddening estimates, we counted the number of stars in the boxes
(``usable stars'') on the CMDs; the magnitude cuts are conservatively
representative of the range within which our monitored stars have
photometric rms of less than 1\% (see Fig. \ref{rms}). For NGC 6253,
we applied a color cut such that our estimate does not include the
evolved background sequence visible in the CMD to the red side of the
box at $V-I \sim 1.4$, since transiting planets around evolved stars
are not detectable due to the large radius of the parent stars. For
NGC 6134, we do not see an evolved background sequence and thus
increased the color range to $V-I \sim 2.5$, in part because the
foreground reddening estimate is higher for this cluster (see Table
\ref{potential}). We note that saturation of our stars sets in at $I
\sim 14.5$. For details, see \S \ref{final_selection}. Estimates for
``usable'' stars for our other OC targets are given in Table
\ref{potential}.
\label{final_target_selection}}
\end{figure}

We also do not change targets during the course of an observing run of
$\sim$20 nights or less (see Fig. \ref{visi}). The justification for
this strategy is simple: to maximize $P_{vis}$.  From Panel b of
Fig. \ref{visi}, one can see that the typical values for $<P_{vis}>$
($P_{vis}$ averaged over all periods between 1 and 5 days) of a $\sim$
20-night observing run with some holes due to weather will reduce the
estimated number of detected planets to 50 -- 70\% of the
``theoretical'' value as calculated in \S \ref{max_number}. Panel c
shows that the efficiency, i.e., how much is added to $<P_{vis}>$ per
night, will peak at around 18 nights for perfect conditions,
justifying our goal of observing every cluster for around 20 nights in
a row.

Alternating cluster targets was suggested by \citet{j96}.
\citet{shl03} adopted an alternating cluster strategy, and while their
detection algorithm could find transits, they found that having only 4
to 6 data points observed during transit was a limiting factor in both
the detection S/N and in discriminating against false positives.
Furthermore, while alternating cluster targets may provide more
monitored stars, this strategy will favor only the 1- to 2-day period
planets if the observing run is not long enough (Figure \ref{visi2}).

Targeting only one cluster during the night further allows us to keep
the stars in the targets OCs on our images at exactly the same place
on the chip (to within less than 1 arcsec). This helps us simplify the
photometry pipeline. In addition, cosmetic problems with the CCD, such
as bad columns or bad pixels, will eliminate the same stars in every
exposure.

%Not correct since variation of +/- 5 pixels:
%Furthermore, keeping objects located at the same positions
%throughout the observing run will reduce scatter in the relative
%photometry caused by small pixel-to-pixel variations across the chip.

%---------------------------------------------------------------------------

\subsection{Dynamic Observing and Optimization of Available
Telescope Time}

We use a real-time approach to maximizing $P_{vis}$ if the allocated
observing time is significantly larger than 20 nights, e.g., $\geq$ 30
nights, based on detecting a single, full transit.

Panel b of Fig. \ref{visi} illustrates that the probability of
detecting an existing {\it single} transit (dashed line) will reach
about 65--70\% after around 10 nights of continuous observing with
10.8 hours per night. As our data reduction pipeline allows us to do
practically real-time data reduction, we can inspect our
highest-quality light curves for the existence of a single transit
after around 10 nights. If, at that point, we do not see any
indication of a single transit anywhere in our data, we will move on
to the next target and observe it for the remainder of the allocated
time. This approach is essentially a comparison of probabilities: the
probability of detecting two transits in a new cluster in the
remaining observing time versus the probability (given no transits
observed so far) of detecting two transits in the current cluster if
we monitor it for the rest of the available observing time.

%If the amount of allocated observing time is $\leq$ 20 nights, then we
%will spend all available time on a single cluster. We aim to spend
%approximately 20 nights on a given OC which is only slightly higher
%than the maximum of the observing efficiency (panel c in
%Fig. \ref{visi}), and is supposed to compensate for the loss of
%observing time due to weather or other factors. Panel c shows that the
%chances for detecting existing planetary transits will monotonically
%increase with the amount of time spent observing this cluster (Panel
%b)--but most efficient for 20 nights.

%---------------------------------------------------------------------------

\subsection{Different Observing Strategies}\label{obs_strategies}

In this Section, we describe how different arrangements of observing
nights affect $P_{vis}$.

At private observatories (such as LCO), different longer-term projects
requiring many nights may compete for time at smaller telescopes such
that their allocation of nights needs to be split.  We explain below
how different ways of dividing observing time between our project and
others affects our likelihood of detecting existing planetary
transits.

In Fig. \ref{comparison}, we illustrate the efficiency of a number of
different observing strategies which may result from such split-time
arrangements. The solid line in all four panels corresponds to
$P_{vis}$ (2 transits detected) of an observing run of 20
uninterrupted nights with 10.8 hours of observing each night.

In Panel a, the dotted line corresponds to $P_{vis}$ of an observing
run spread over 40 nights (10.8 hours per night), during which we
observe only for the first two nights out of every four. $P_{vis}$ is
approximately the same as the one for 20 consecutive nights. The
$P_{vis}$ averaged over all periods (1 day -- 5 days), $<P_{vis}>$, of
the 20-consecutive-nights observing run is 0.681. The same $<P_{vis}>$
for the 2-nights-on, 2-nights-off strategy over 40 nights is 0.666.
We note that the 2-on, 2-off strategy may impose difficulties in (1)
the period determination due to aliasing effects (see below), and (2)
the loss of observing time per night due to the drift of the sidereal
time over the course of such a long observing run.

In Panel b, the dotted line showcases the result of observing only the
first half of every night for 40 nights in a row. The likelihood of
detecting existing transits is reduced significantly ($<P_{vis}> \sim$
0.437).  For a strategy of observing a third of every night for 60
nights, as shown by the dotted line in Panel c, $<P_{vis}>$ goes down
to 0.007.

Note that none of these numbers takes into account the drift of the
sidereal time which would reduce the number of hours of observability
during the night as a function of declination of the target. As a
result of the sidereal drift, $<P_{vis}>$ would be reduced from 0.748
for a run of 20 consecutive nights to 0.705 for a run of 40 nights
with the 2-nights-on, 2-nights-off strategy for NGC 6208, assuming it
is perfectly centered in RA at the midpoint of the hypothetical
observing run. We calculated similar decreases (on the order of 5\% or
less) in $<P_{vis}>$ when comparing the two observing strategies for
the other clusters in Table \ref{potential}.

Finally, Panel d illustrates the aliasing effect of only observing 2
out of 4 nights. The dotted line in Panel d corresponds to the
probability of detecting two existing transits from which the period
can be correctly determined when applying the 2-nights-on,
2-nights-off strategy over the course of 40 nights. $<P_{vis}>$ of the
dotted line is 0.356, meaning that only about half (0.356/0.666) of
all transit observations would result in a correct calculation of the
period, whereas the rest would suffer from aliasing effects. Note that
this ratio is sensitively dependent on the period itself, as
illustrated by the dotted line. For comparison, a strategy of
1-night-on, 1-night-off would produce $<P_{vis}>$ (two transits
observed) of 0.677, but a $<P_{vis}>$ (no aliasing) of only 0.286,
meaning that a larger fraction (1-(0.286/0.677)$\simeq 58$\%) of
observed transits would result in an incorrect calculation of the
period. The strategy of continuously observing for 20 nights will give
a $<P_{vis}>$ (no aliasing) of 0.406, i.e., an correct estimate of the
period in 0.406/0.682$\sim$60\% of the cases.

We thus conclude that while having 20 consecutive, uninterrupted
nights is clearly the most favorable solution, we can tolerate the
strategy where we observe 2 out of every four nights without a
significant loss in $<P_{vis}>$, but which will increase the
probability of aliasing effects in the period determination.

%---------------------------------------------------------------------------

\subsection{Contamination by Galactic Field Stars}\label{contamination}

Estimates of background or foreground stellar contamination to OCs are
valuable since they are the basis upon which statistical estimates of
planet frequency among OC members are based, regardless of whether a
planet was detected or not. In order to get a handle on contamination,
we observe two control fields per target cluster at the same Galactic
latitude, approximately a degree away from the OC. These observations
are ideally taken in $V$ and $I$, using the same exposure time as for
the cluster field, and taken in the same weather and seeing
conditions. To first order, the excess number of stars in the cluster
field will be representative of the number of cluster members,
subject, of course, to uncertainty due to fluctuations of background
and foreground star counts.

Figures \ref{n2660_histogram} and
%, \ref{n2660_positions}, and
\ref{n2660_cmds} show this approach for estimating contamination for
the observed OC NGC 2660.  Fig. \ref{n2660_histogram} compares the
stellar density (measured in units of stars per 100 pix $\times$ 100
pix on the CCD with $13.0 < I < 17.0$) as a function of radial
distance from the CCD center of the cluster image of NGC 2660 (solid
line) and two control fields (dotted and dashed lines) at the same
Galactic latitude offset by 1 degree in the sky in either direction.
%The cluster field shows the concentration of stars towards
%the middle of the CCD in Fig. \ref{n2660_positions} where we show the
%locations of the stars with $13.0 < I < 17.0$ for the cluster (left
%panel) and both control fields (middle and right panel).  
The comparison between the CMDs of the cluster and control fields is
shown in Fig. \ref{n2660_cmds}. Although the cluster main sequence is
not clearly visible in its CMD, one may nevertheless see a higher
density of stars with respect to the control field CMDs at colors
red-ward of $V-I \sim 1.2$, as well as a red clump at around $I \sim
13$ and $V-I \sim 1.4$.  The total number of stars within $13.0 < I <
17.0$ in the cluster field is around 3500 stars versus 2700 and 2900
stars in the two control fields. The contamination over the entire CCD
field is thus around 80\%, and approximately 30\% towards the center
of the field out to a distance of around 4 arcmin.
%, in perfect agreement with
%the study by \citet{nsp02}.

Figures \ref{n6208_histogram} and
%\ref{n6208_positions}, and
\ref{n6208_cmds} illustrate how much more severe this contamination
can be, using NGC 6208 as an example (for which we only have data for
a single control field). Fig. \ref{n6208_histogram} compares the
stellar density (same units as Fig. \ref{n2660_histogram}) as a
function of radial distance from the CCD center of the cluster image
of NGC 6208 and a control field at the same Galactic latitude offset
by 1 degree in the sky. Here, the cluster excess stars do not seem to
be very centrally concentrated (cf. Fig. \ref{n2660_histogram}).
% which can also be seen
%in Fig. \ref{n6208_positions} where we show the positions of the stars
%with $13.0 < I < 17.0$ in both the cluster and control field.
Finally, the comparison between the CMDs of the cluster and control
field show a slight excess of stars in the cluster CMD at bright
magnitudes (Fig. \ref{n6208_cmds}). These excess stars (located around
$I \sim 13.0, V-I \sim 0.7$) are evenly distributed over the cluster
field, and are approaching the bright limit of our photometry (see
Fig. \ref{rms}). The total number of stars in the above magnitude
range in the cluster field is around 6200 stars versus 6000 stars in
the control field. This would amount to a contamination of 97\% over
the entire field, and of around 85\% in the inner 5 arcmin. This heavy
contamination and the associated high density of the region in which
NGC 6208 is located was noticed by \citet{l72} and reiterated in
\citet{pm01}. This cluster contamination of 97\% is similar to the
94\% contamination of NGC 6819 estimated by \citet{shl03}. Taking into
account this high rate of contamination, only a few hundred stars of
high relative photometric precision are actually cluster members.
Detecting transits around field stars, however, is still useful in a
number of ways, outlined in \S \ref{transits}.

%---------------------------------------------------------------------------

%\subsection{Cluster Parameters}\label{parameters}
%
%In order to determine or constrain cluster parameters such as age,
%metallicity, and reddening, we obtain $BVRI$ color sequences.  These
%color sequences help put further constraints on contamination
%estimates based on assigning membership probabilities of individual
%stars as a function of their position with respect to the cluster main
%sequence in the various multi-color CMDs and color-color diagrams, as
%suggested by, e.g., \citet{shl03,bhb03}.
%
%Importantly, we also include short exposures to complete the CMD at
%brighter magnitudes than our target stars to permit isochrone-fitting
%for the determination of the astrophysical OC parameters.

%---------------------------------------------------------------------------

\section{EXPLORE/OC Photometric Data Reduction Methods and 
Spectroscopy Follow-Up}\label{photometry}

%---------------------------------------------------------------------------

The EXPLORE/OC strategies concerning photometric data reduction and
spectroscopy follow-up work are described here in a brief, preliminary
way. More detailed descriptions will follow along with the
presentations of our results of the individual OCs.

\subsection{Photometry Data Reduction Pipeline}

After the standard IRAF\footnote{IRAF is distributed by the National
Optical Astronomy Observatories, which are operated by the Association
of Universities for Research in Astronomy, Inc., under cooperative
agreement with the NSF.} image-processing routines, our stellar
photometry for the reduction of individual images is performed by an
algorithm which will be described in detail in an upcoming publication
(Yee et al. 2004, in preparation), and is outlined in principle in
\S 4.3 of M03. We will only provide a very brief overview here.

At the heart of our aperture photometry algorithm is the accurate
placement of the aperture relative to the centroid of the star under
investigation.  This is an important issue due to the relative
brightness of the sky with respect to the monitored stars.  To
minimize the contribution of sky noise and other systematics, we use a
relatively small aperture (2--3 seeing disks), which further improves
photometry in the situation of moderate crowding (with star
separations of a few seeing disks; see \S \ref{blending}).  To
achieve the accurate placement of the aperture crucial for obtaining
high-precision relative photometry, we use an iterative sinc-shifting
technique to re-sample every star individually such that the central 3
$\times$ 3 pixels are symmetrically located about the centroid of the
respective star's PSF.  Performing this shift for every object in the
frame is then equivalent to using an identical placement of the
aperture masks for every object, ensuring proper relative photometry.
With such re-sampling, aperture photometry of different aperture radii
can be performed simply by using integer pixel masks of various sizes.
Sinc-function re-sampling is an ideal method for shifting an image
that is Nyquist sampled since it preserves resolution, noise
characteristics, and flux \citep{h77,y88}.

Our relative photometry is then performed by iteratively determining
the most stable stars within subregions of the CCD field. All other
stars within the same subregion are shifted to the photometric system
of these reference stars, thereby using iterations to minimize the
scatter and to remove outliers from the calculation of the photometric
shift. The number of iterations, criteria for outlier removal, size of
the subregions, and minimum number of stars per subregion are
parameters that vary for each dataset. Some of the light curves
produced by this algorithm are shown as examples in Fig.
\ref{lightcurves} and illustrate our potential to detect 1\% amplitude
signals within the intrinsic scatter of the high-precision photometry
for the target magnitude range.

%---------------------------------------------------------------------------

\subsection{Spectral Type Determination Follow-Up}
\label{spectroscopy}

\begin{figure}
\epsscale{0.7} \plotone{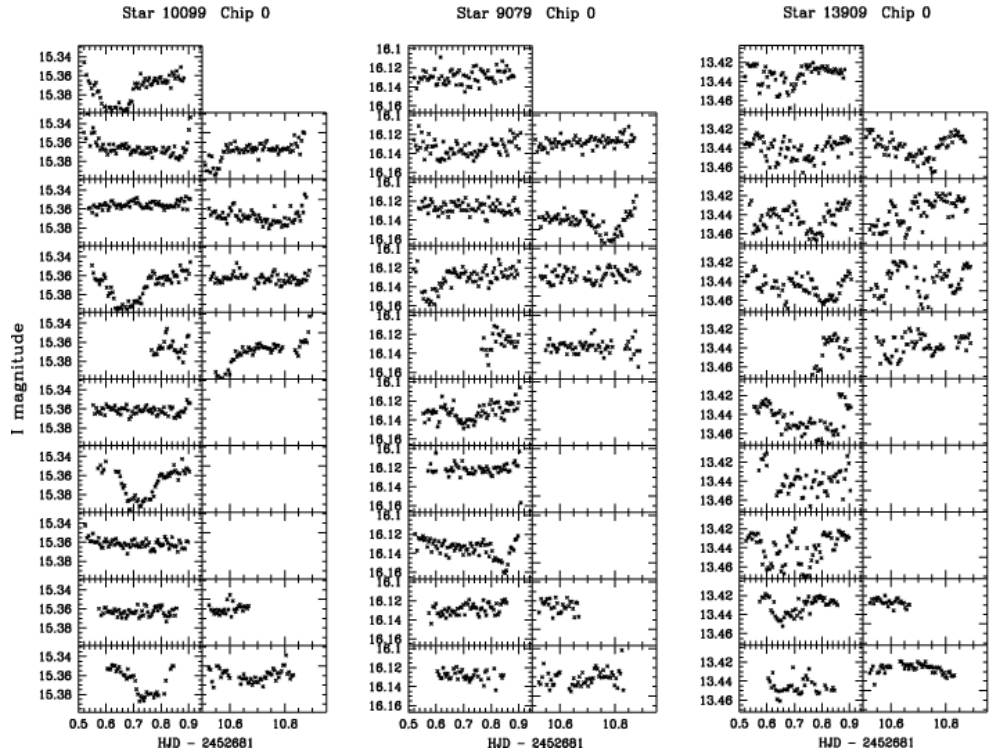}
\caption{Examples of light curves derived from our high-precision
relative photometry (see \S \ref{photometry}) of our NGC 2660
data. Every panel of one of the light curves represents the data taken
during a single night, starting with night 1 on the bottom left.
Night 2's data are shown in the panel directly above it, night 3 above
that and so on. No data were obtained during nights 13--15 due to
telescope scheduling, and nights 6 and 12 were only partially useful
due to weather. All three displayed light curves show the
low-amplitude, transit-like signal we are looking for in our
survey. They are, however, most likely caused by non-planetary
phenomena such as a larger-sized companion (left panel) or grazing
binaries (middle and right panels). Our preliminary work on
spectral type determination indicates that star 10099 (left panel) is
an early G star, star 9079 (middle) is a late A star, and star 13909
(right) is somewhere between F2 and F5. \label{lightcurves}}
\end{figure}

\begin{figure}
\epsfig{file=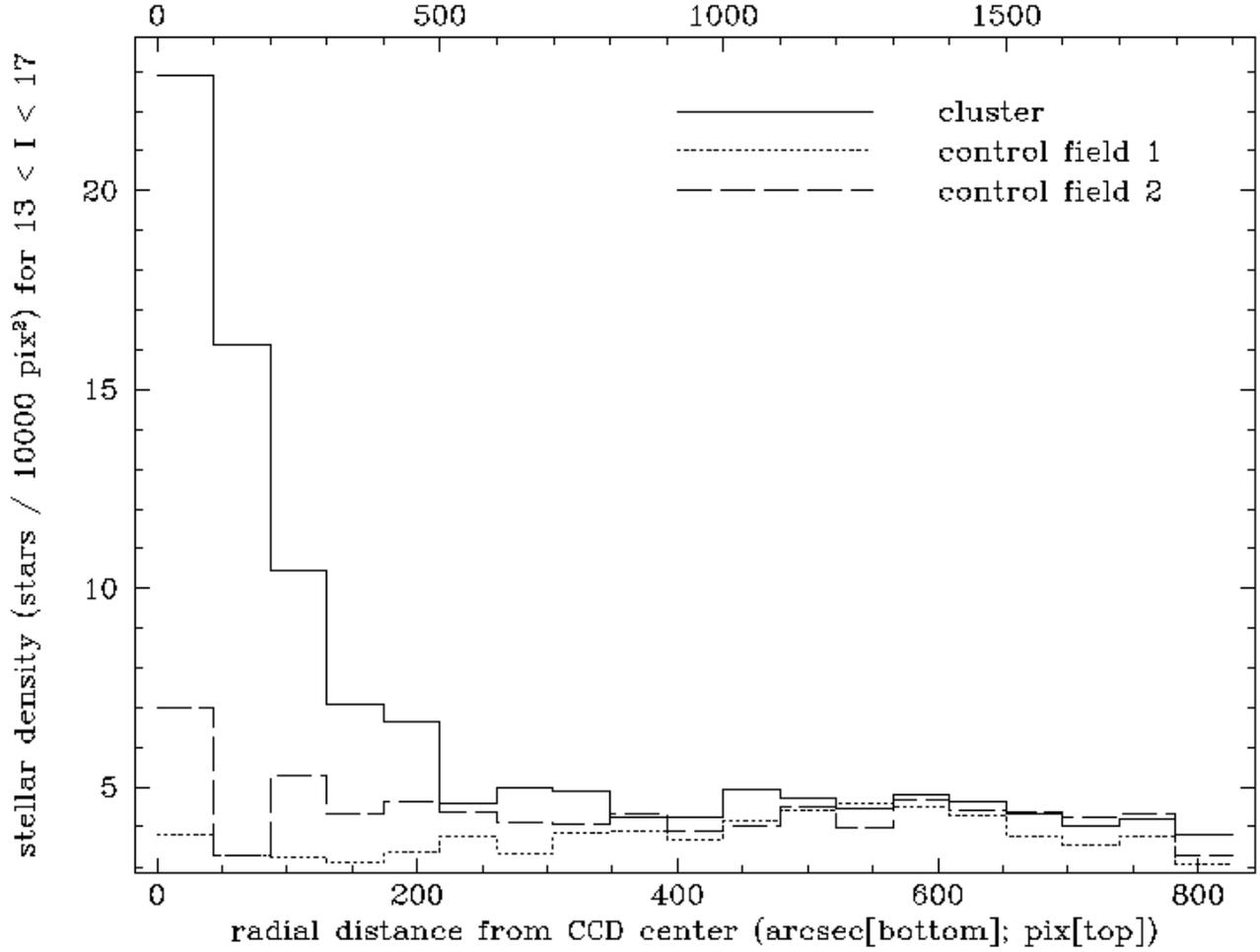, angle=-90, width=\linewidth}
\caption{\label{n2660_histogram} This figure compares the stellar
density (measured in stars per 100 pix $\times$ 100 pix on the CCD
with $13.0 < I < 17.0$) as a function of radial distance from the CCD
center of the NGC 2660 open cluster image (solid line) and two control
fields (dotted and dashed lines) at the same Galactic latitude offset
by 1 degree in the sky in either direction in Galactic longitude. The
contamination is around 80\% over the whole field of the CCD, and
approximately 30\% for the inner $\sim$ 4 arcmin. For details, see \S
\ref{contamination}.}
\end{figure}

\begin{figure}
\plotthree{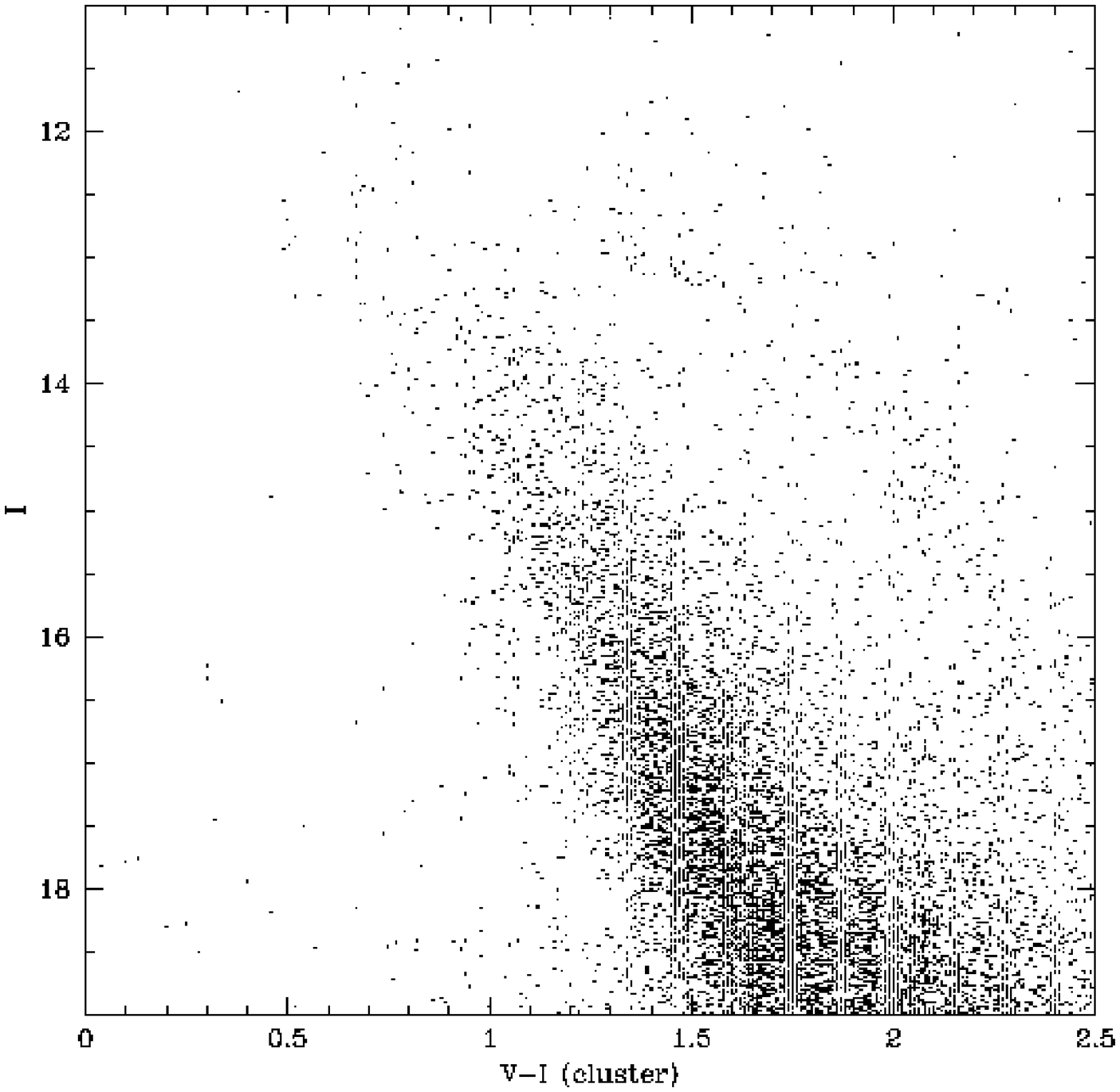}{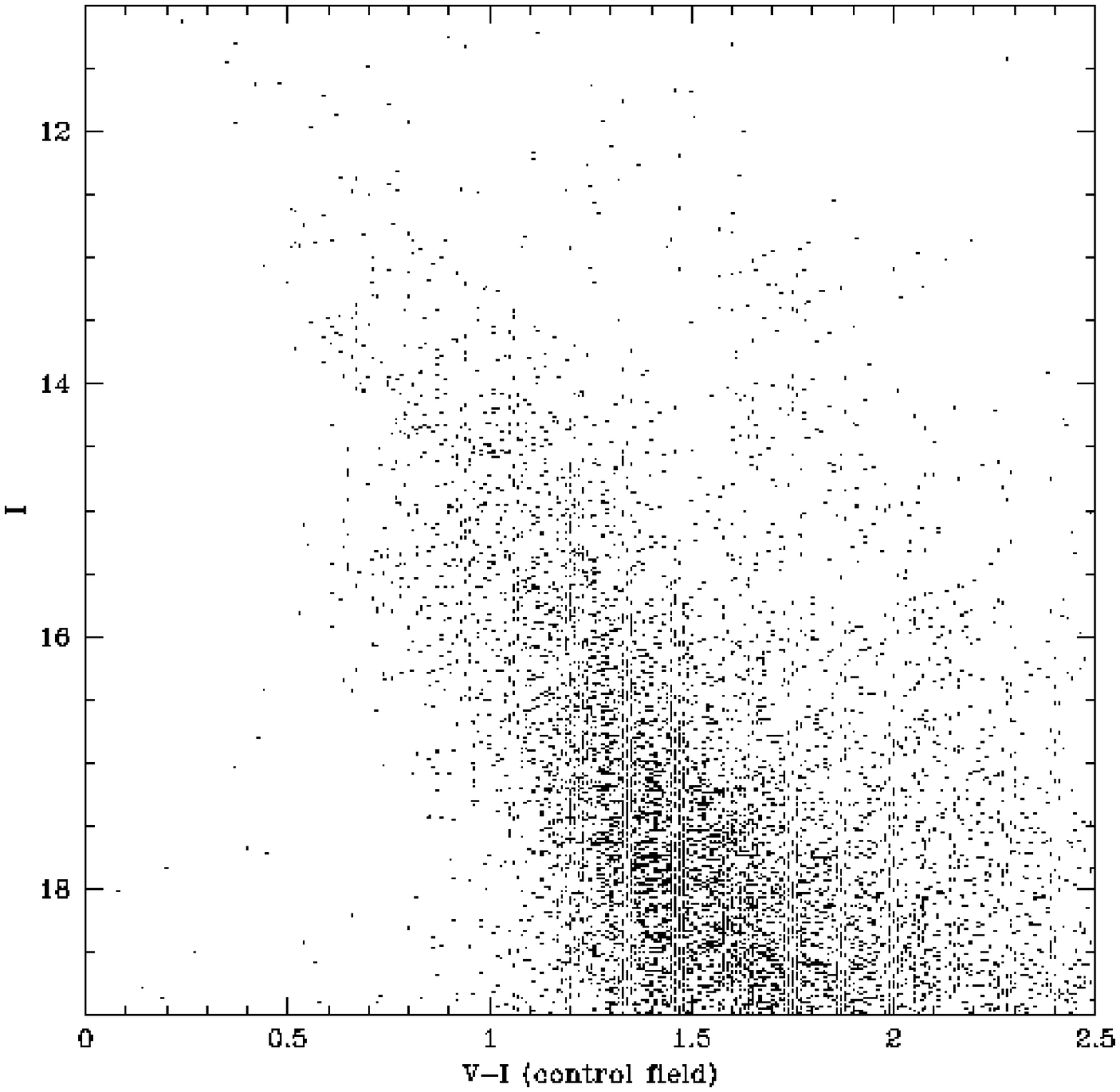}{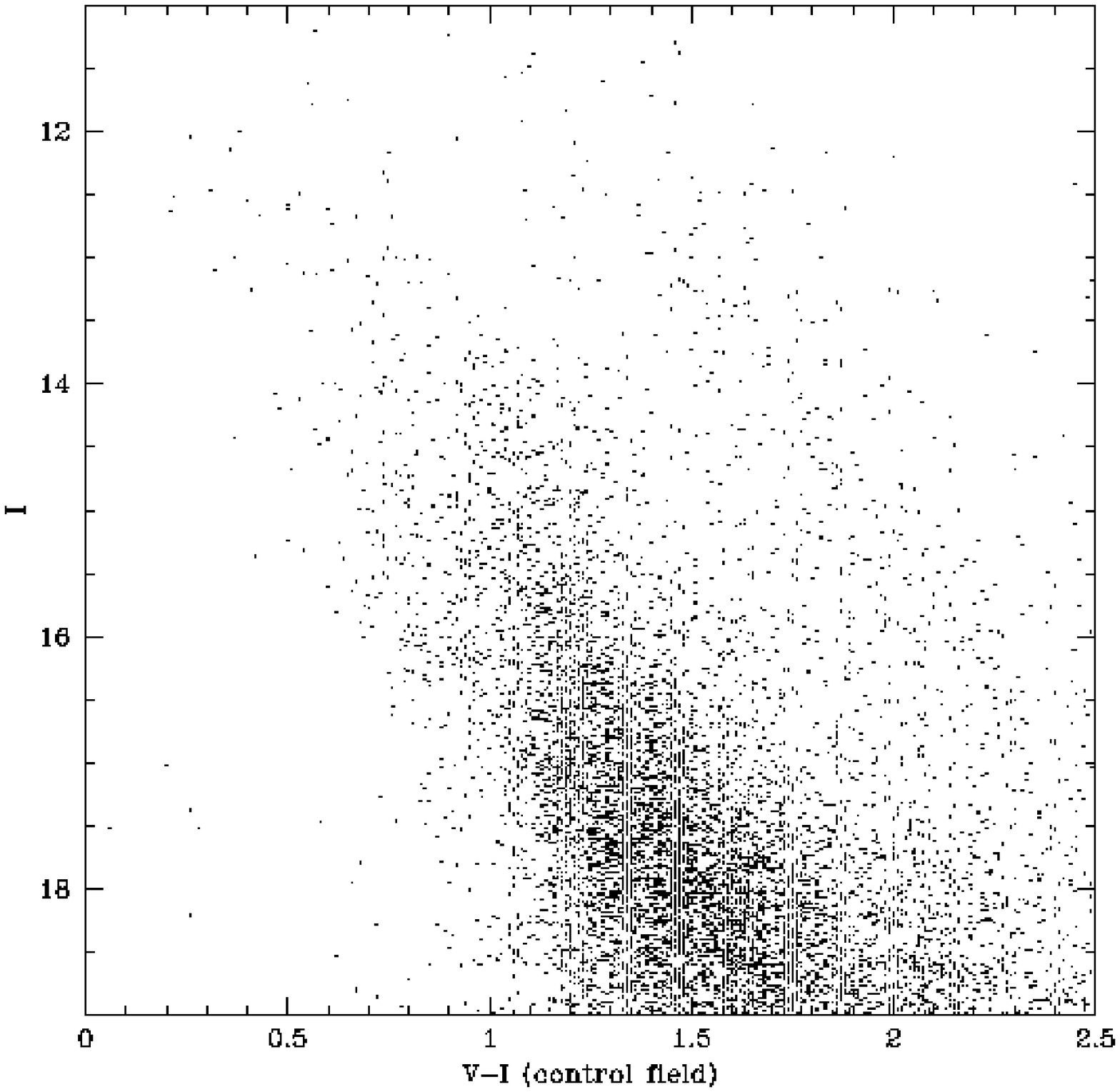}
\caption{\label{n2660_cmds} This figure shows the CMDs of the field
centered on NGC 2660 (left panel) and two control field (middle and
right panel) at the same Galactic latitude offset by 1 degree in the
sky. }
\end{figure}

\begin{figure}
\epsfig{file=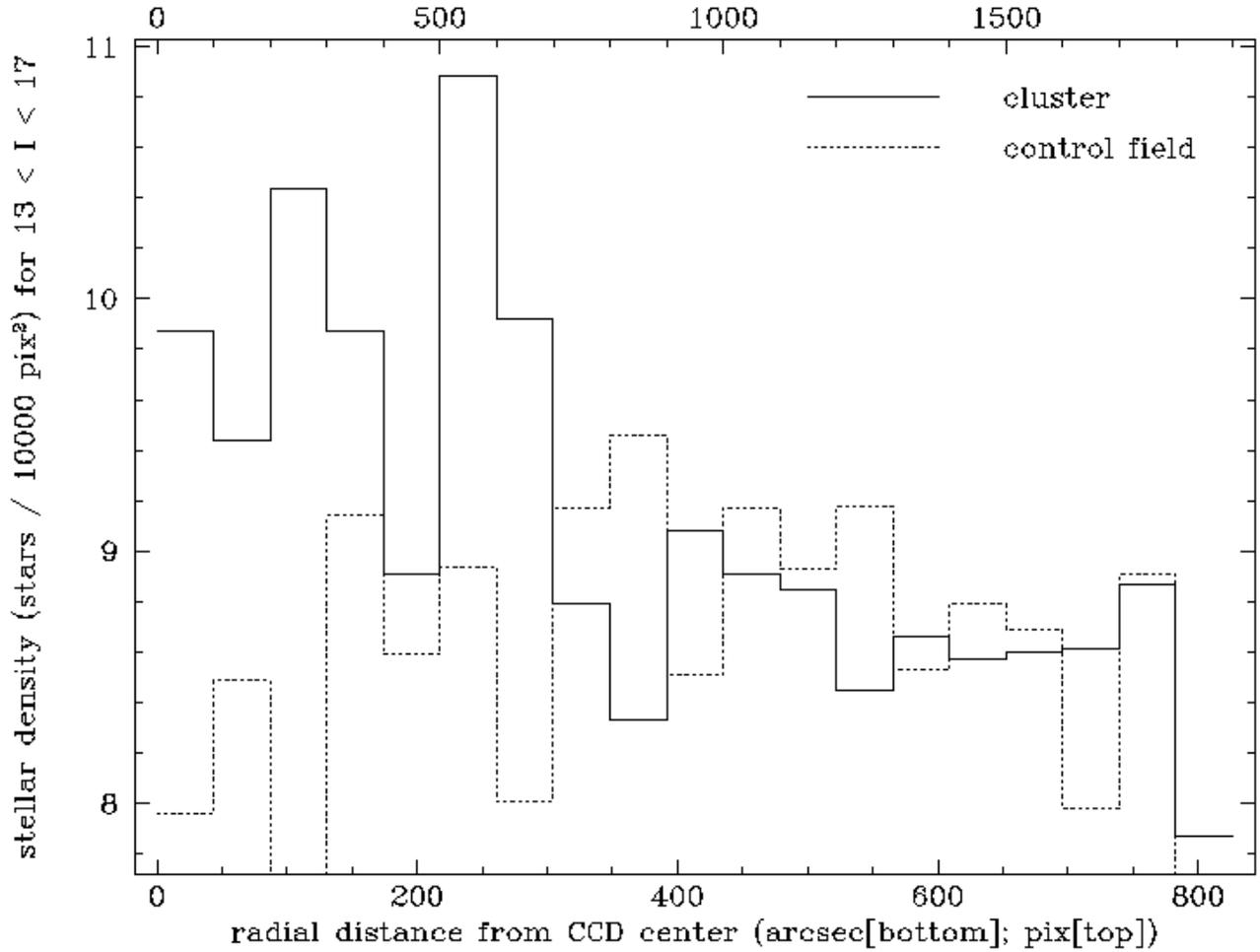, angle=-90, width=\linewidth}
\caption{\label{n6208_histogram} This figure compares the stellar
density (measured in stars per 100 $\times$ 100 pix on the CCD with
$13.0 < I < 17.0$) as a function of radial distance from the CCD
center of the NGC 6208 open cluster image (solid line) and a control
field (dotted line) at the same Galactic latitude offset by 1 degree
in the sky. The contamination is around 97\% when integrated over the
entire CCD, and around 85\% towards the inner 5 arcmin. For details,
see \S \ref{contamination}.}
\end{figure}

\begin{figure}
\plottwo{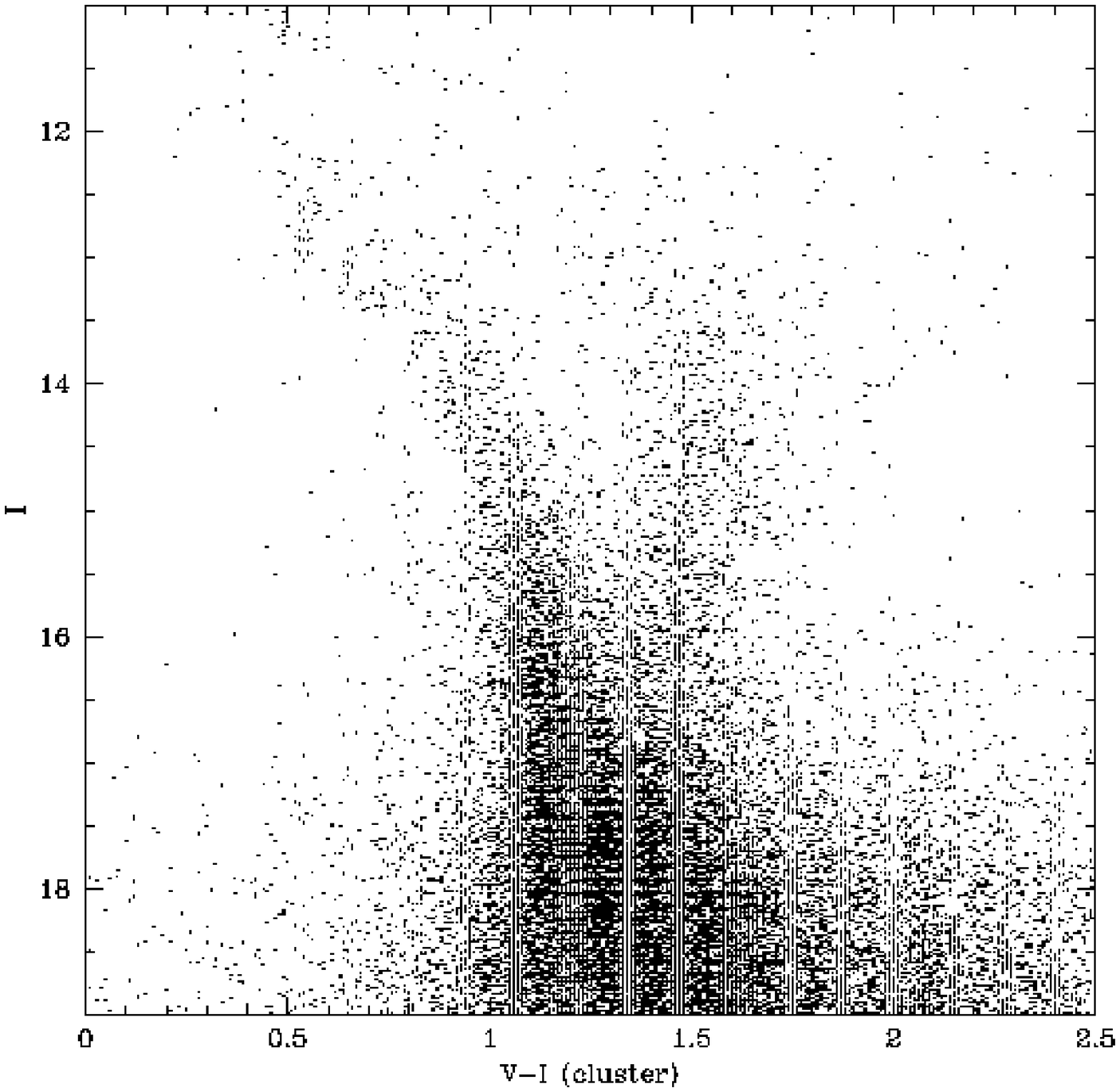}{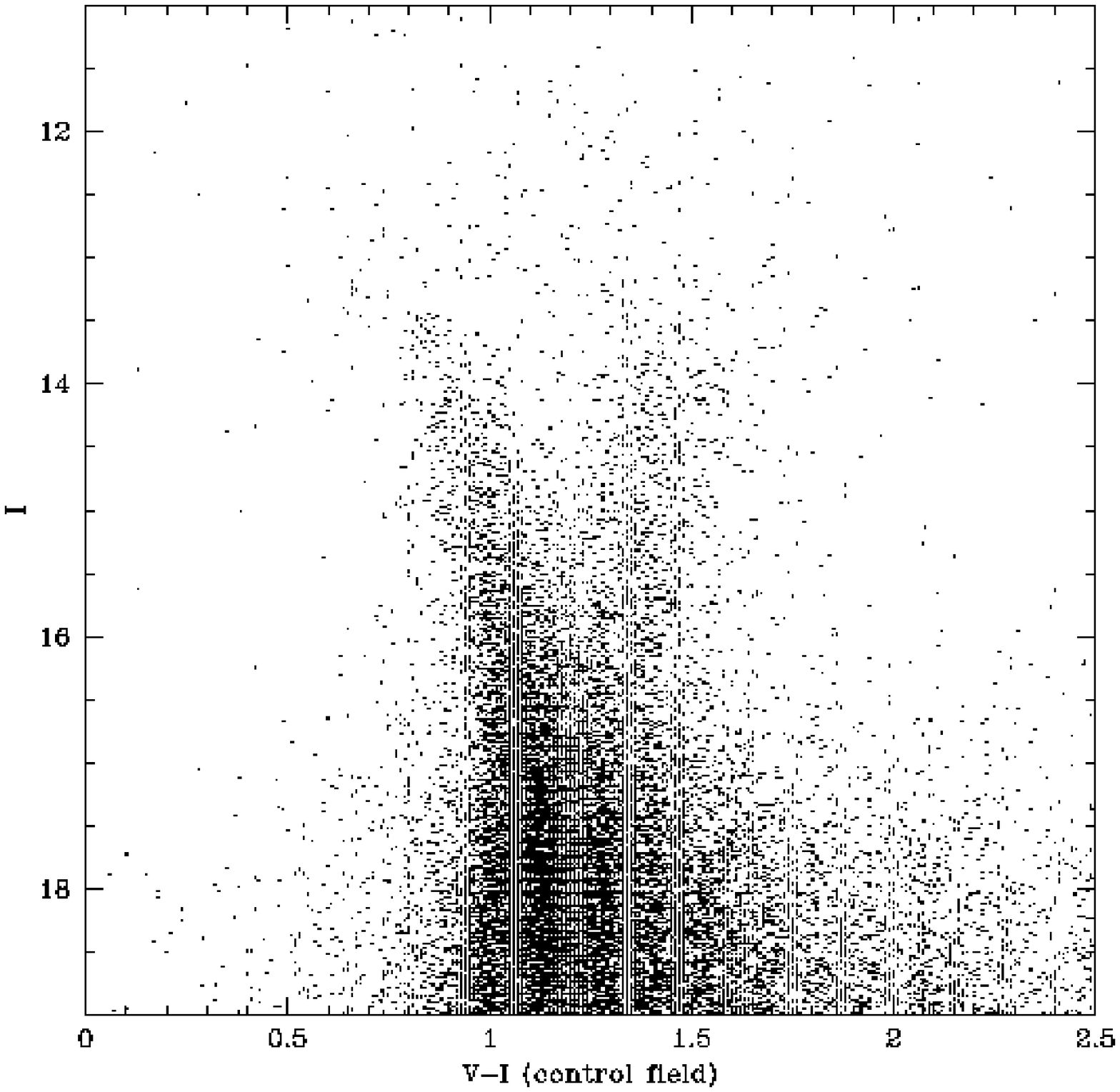}
\caption{\label{n6208_cmds} This figure shows the CMDs of the field
centered on NGC 6208 (left panel) and a control field (right panel) at
the same Galactic latitude offset by 1 degree in the sky. The
comparison between the CMDs shows a slight excess of stars in the
cluster CMD at bright magnitudes. These excess stars (located around
$I \sim 13.0, V-I \sim 0.7$) are evenly distributed over the cluster
field, and are approaching the bright limit of our photometry (see
Fig. \ref{rms}). Previous studies \citep{l72,pm01} already mentioned
the difficulty in separating the cluster main sequence from the Galactic
disk population. For details, see \S \ref{contamination}.}
\end{figure}

We determine spectral types for our planet candidate stars to provide
an independent measure of their sizes which may help break
degeneracies in the photometric solution such as period aliasing or
stellar blends, and may thus determine whether or not costly RV
follow-up work is desired \citep{sm03,tks04b,tks04a}.  We have
obtained spectral data using a variety of instruments which include
the Boller \& Chivens Spectrograph and the IMACS Multi-Object Imaging
Spectrograph (both on the Magellan 6.5m Telescopes), as well as the
Wide-Field Re-imaging CCD in Grism/Multi-slit mode on the LCO du Pont
2.5m Telescope. We are currently analyzing spectral data for our
potential candidates (examples in Fig. \ref{lightcurves}) from our
work on NGC 2660 and NGC 6208 to determine the exact nature of each of
the systems. Preliminary results are given in Fig. \ref{lightcurves}.

Spectral type determination of non-planet-candidate stars in the field
will give estimates of the foreground reddening along the line of
sight and differential reddening across the field, and provide an
independent check on the determination of cluster distance by
isochrone fitting.  Furthermore, the knowledge of the spectral types
of a representative set of stars (tens or hundreds of stars) will
provide an additional means of estimating contamination of the sample
by Galactic field stars and will allow us to determine the parent
sample of non-cluster members. 

%---------------------------------------------------------------------------

\section{Summary}\label{future}

Open clusters are regarded as suitable planet transit monitoring
targets because they represent a relatively large number of coeval
stars of the same metallicity located at the same distance (\S
\ref{transits_ocs}). Four groups are now monitoring over a dozen open
clusters for short-period transiting planets (see \S \ref{transits}).

We reviewed the main challenges facing transit searches (\S
\ref{mainchallengeTS} and OC surveys in particular \S
\ref{mainchallengeOC}). In addition to the difficulties involved in
any transit search, they include:

\begin{itemize}

\item
The relatively low number of stars at high relative photometric
precision (1--1.5\%) compared to Galactic field surveys of roughly the
same magnitude range: $\sim$ 5,000 compared to $\sim$ 50,000 stars
respectively (though the difference in field size is not taken into
account here). This number is similar to the number of stars obtained
by the 6$\times$6 degree$^{2}$ shallow transit surveys of brighter
stars.

\item
The severe contamination by Galactic field stars, up to 97\% in our
clusters for stars at $13 < I < 17$.

\item
Differential reddening may be problematic in fitting isochrones.

\end{itemize}

Just like for field transit surveys, OC transit surveys need
to maximize the number of stars with high photometric precision,
maximize the probability to detect an existing transit, and not be
swamped by false positive transit signals.

We presented aspects of the EXPLORE/OC planet transit survey design
that were considered to meet some of the major challenges facing
transit surveys. Target selection is a key aspect to survey design
with the number of factors over which to optimize (richness,
observability, age, distance, foreground reddening) actually limiting
the number of available targets for a given observing time and
Galactic location.  We choose high-cadence observing in order to
sample transits well enough to easily rule out false positives such as
grazing eclipsing binaries, and to use the unique solution method
\citep{sm03} to estimate planet and star parameters. We have shown
that with an adopted exposure time and a given telescope, the distance
of the cluster can be chosen to target certain spectral types. For the
EXPLORE/OC project we do not alternate OCs in a given observing run
but instead remain on one cluster in order to maximize finding planet
transits. We have shown that this strategy optimizes the probability
to detect an existing transiting planet with periods of 2 -- 5 days
days if observing runs are around 20 days. The single cluster
approach, together with near real-time data reduction, allows us to
use our dynamic observing strategy for long observing runs ($>$ 30
nights): if a single full transit is not seen within 10 days, the
strategy is to move on to another cluster.

EXPLORE/OC is the only OC planet transit survey operating in the
southern hemisphere.  We have presented some preliminary data on the
OCs NGC 2660 and NGC 6208 in order to illustrate the main challenges
facing cluster surveys as well as to illustrate our survey design
strategy.  Our $I$-band, high-cadence photometric monitoring with the
LCO 1m Telescope typically attains 1\% precision in our relative
photometry for around 3000--5000 stars per OC field in the range $14.5
< I < 17$ with 5 min exposures. For a cluster at a distance of 1 kpc
and $E_{B-V}$ of 0.2, this magnitude range corresponds to a range of
spectral types between mid-G to early M. We have obtained data on
three additional open clusters: IC 2714, NGC 5316, and NGC 6253, and
plan to target 4--5 more clusters.

With the $\sim$12 OCs currently being monitored and analyzed by the
four existing OC surveys, there is a good chance that some
short-period planets will be detected in the near future.  Because of
the potentially large contamination, and poor availability of physical
data on many clusters in the literature, any detected planets should
individually be confirmed as cluster members. Furthermore,
characterizing of the cluster parameters is important \citep{bgd04}.
With a limited number of stars per cluster, severe contamination from
field stars, and considering the finite magnitude range for which
high-precision photometry can be obtained, only several hundred to a
few thousand cluster members are monitored with high enough
photometric precision to detect planet transits; nevertheless, planet
transits detected in the contaminating field stars are also useful. If
one optimizes the important selection criteria, partly due to the
paucity of old clusters, most of the suitable OCs for photometric
planet searches and radial-velocity follow up can be searched with a
reasonable amount of time and effort.

%---------------------------------------------------------------------------

\acknowledgements{ }

We would like to thank Ted von Hippel for very helpful insights on the
problem of Galactic contamination issues and on the selection of
observing targets. Furthermore, we thank the anonymous referee,
B. Scott Gaudi for his insightful comments on this manuscript, and
Marten van Kerkwijk for useful discussions. Finally, we extend our
gratitude to the staff at Las Campanas Observatory for their unmatched
dedication to optimizing everything about our observing runs and time
spent at LCO. This work was in part supported by NSF grant
AST-0206278.  BLL was supported by a NSERC Graduate Fellowship and a
Walter C. Sumner Memorial Fellowship. GMO was supported by a Clay
Fellowship at the Smithsonian Astrophysical Observatory.

%---------------------------------------------------------------------------
\bibliography{planets}
%---------------------------------------------------------------------------
% Tables
%---------------------------------------------------------------------------
%\clearpage

% Table 1 - deluxetable format - enclosed
\begin{deluxetable}{rrccrrrrrccc}
\tabletypesize{\scriptsize}
\tablecaption{Potential Open Cluster Targets \label{potential}}
\tablewidth{0pt}
\tablehead{
\colhead{Cluster} & \colhead{$D$ (pc)} & \colhead{$E_{B-V}$} & 
\colhead{$M_{I_{limit}}$\tablenotemark{a}} & \colhead{$\alpha_{2000}$} & 
\colhead{$\delta_{2000}$} & \colhead{l} & \colhead{b} &
\colhead{[Fe/H]} & \colhead{log(age)} & 
\colhead{richness\tablenotemark{b}} &
\colhead{1\%-rms stars\tablenotemark{c}}
}
\startdata
NGC 2423 &766 &0.097 &7.39 &07 37 06.7 &--13 52 17 &230.5 &   3.5 & +0.14  &8.867 & 4 & 1400\\
NGC 2437 &1375&0.154 &6.01 &07 41 46.8 &--14 48 36 &231.9 &   4.1 & +0.06  &8.390 &  & 1600\\
NGC 2447 &1037&0.046 &6.83 &07 44 29.2 &--23 51 11 &240.0 &   0.1 & +0.03  &8.588 & 4 & 1900\\
NGC 2482 &1343&0.093 &6.18 &07 55 10.3 &--24 15 17 &241.6 &   2.0 & +0.12  &8.604 & 2 & \\
NGC 2539 &1363&0.082 &6.17 &08 10 36.9 &--12 49 14 &233.7 &  11.1 & +0.14  &8.570 &  &\\
NGC 2546 &919 &0.134 &6.92 &08 12 15.6 &--37 35 40 &254.9 & --2.0 & +0.12  &7.874 & 3 & 1900\\
NGC 2571 &1342&0.137 &6.09 &08 18 56.3 &--29 44 57 &249.1 &   3.6 & +0.08  &7.488 &  &\\
NGC 2660\tablenotemark{d} &2826&0.313 &4.14 &08 42 38.0 &--47 12 00 &265.9 & --3.0 & --0.18  &9.033 & 5 & 2750 \\
 IC 2488 &1134&0.231 &6.28 &09 27 38.2 &--57 00 25 &277.8 & --4.4 & +0.10  &8.113 &  & 1600\\
NGC 3114 &911 &0.069 &7.07 &10 02 29.5 &--60 07 50 &283.3 & --3.9 & +0.02  &8.093 & 2 & 2900\\
 IC 2714 &1238&0.341 &5.87 &11 17 27.3 &--62 43 30 &292.4 & --1.8 & --0.01  &8.542  &  & 2750\\
NGC 5316 &1215&0.267 &6.06 &13 53 57.2 &--61 52 00 &310.2 &   0.1 & +0.13  &8.202  & 3 & 2800 \\
NGC 5822 &917 &0.150 &6.90 &15 04 21.2 &--54 23 47 &321.6 &   3.6 & --0.03  &8.821 & 4 & 2600 \\
NGC 6025 &756 &0.159 &7.30 &16 03 17.7 &--60 25 53 &324.6 & --5.9 & +0.23  &7.889  & 3 &\\
NGC 6067 &1417&0.380 &5.51 &16 13 11.0 &--54 13 08 &329.7 & --2.2 & +0.14  &8.076  &  &\\
NGC 6087 &891 &0.175 &6.91 &16 18 50.5 &--57 56 04 &327.7 & --5.4 & --0.01  &7.976 & 3 &\\
NGC 6134 &913 &0.395 &6.43 &16 27 46.5 &--49 09 04 &334.9 & --0.2 & +0.18  &8.968  & 4 & 2850 \\
NGC 6208\tablenotemark{d} &939 &0.210 &6.73 &16 49 28.1 &--53 43 42 &333.8 & --5.8 & 0.00  &9.069 & 4 & 3250\\
NGC 6253 &1510&0.200 &5.72 &16 59 05.1 &--52 42 32 &335.5 & --6.3 & +0.36  &9.70  &  & 3400\\
NGC 6259 &1031&0.498 &5.97 &17 00 45.4 &--44 39 18 &342.0 & --1.5 & +0.02  &8.336 &  &\\
 IC 4651 &888 &0.116 &7.03 &17 24 42.0 &--49 57 00 &340.1 & --7.9 & +0.09  &9.057 & 4 &\\
NGC 6425 &778 &0.399 &6.77 &17 47 01.6 &--31 31 46 &357.9 & --1.6 & +0.07  &7.347 & 2 &\\
\enddata

\tablenotetext{a}{Limiting absolute $I$ magnitude to which we can
observe with a photometric precision of 1\% or better for 300s
exposure time at the Swope 1m Telescope obtained during photometric
conditions and good seeing. This value is obtained by conservatively
(cf. \S \ref{max_number}) assuming that the apparent $I_{limit} = 17$,
and that $A_{I} = 1.94 E_{B-V}$ \citep{sfd98}.}

%\tablenotetext{b}{Average observability for $\sim 20$ night observing
%run center on the date in parentheses. We consider a cluster to be
%observable when the airmass is smaller than 2, and the sun's altitude
%is at least $10^{\circ}$ below the horizon. Calculations performed
%with SKYCALC by John Thorstensen.}

\tablenotetext{b}{Richness class as given in \citet{ja82,c00} if
available. Range: 1 (sparse) to 5 (most populous). Should be regarded
as a lower limit to the actual richness of the cluster since it
depends on the depth of the study from which it was derived (see
\S \ref{observing}).}

\tablenotetext{c}{The approximate number of {\it main-sequence stars}
(if available) for which we expect to achieve a relative photometric
precision of 1\% or better for 5-min exposures with the Swope
Telescope (see \S \ref{final_selection} and
Fig. \ref{final_target_selection} for details). Should be regarded as
a lower limit to the number of stars around which we are able to
detect planetary transits (cf. \S \ref{max_number}).}

\tablenotetext{d}{Previously observed cluster; see \citet{blm04} and
\citet{lbm04} for preliminary results on NGC 2660 and NGC 6208,
respectively. We note that NGC 2660 (our first target) was chosen for
its estimated richness and its observability given the allocated
observing time. It turned out to be a non-optimal target due to its
larger distance and correspondingly brighter limiting absolute $I$
magnitude.}

\tablecomments{This table shows our previously observed OCs plus a
number of potential target clusters which we chose based on the
criteria outlined in \S \ref{targets} and \S \ref{targets_exp}. Data
were taken from the WEBDA database; metallicities from \citet{taa97},
available at \url{http://obswww.unige.ch/webda/feh\_twarog.html}. The
column headers are cluster name, distance in parsecs, foreground
reddening, limiting absolute $I$ magnitude, $\alpha$, $\delta$,
Galactic longitude, Galactic latitude, metallicity, logarithm of the
age (in years), the value for the estimated richness class, and the
approximate number of stars in the field with relative photometric 
precision of 1\% or better.}

\end{deluxetable}

%---------------------------------------------------------------------------

%\clearpage

%---------------------------------------------------------------------------
% Figures
%---------------------------------------------------------------------------
%\clearpage

%---------------------------------------------------------------------------

%---------------------------------------------------------------------------

%---------------------------------------------------------------------------

%---------------------------------------------------------------------------

%---------------------------------------------------------------------------

%---------------------------------------------------------------------------

%---------------------------------------------------------------------------

%---------------------------------------------------------------------------

%\begin{figure}
%\epsscale{0.9}
%\plotthree{figures/vonBraun.fig9a.eps}{figures/vonBraun.fig9b.eps}{figures/vonBraun.fig9c.eps}
%\caption{\label{n2660_positions} This figure shows the positions of
%stars ($13.0 < I < 17.0$) in the field centered on NGC 2660 (left
%panel) and two control fields (middle and right panels) at the same
%Galactic latitude offset by 1 degree in the sky.}
%\end{figure}

%---------------------------------------------------------------------------

%---------------------------------------------------------------------------

%---------------------------------------------------------------------------

%\begin{figure}
%\plottwo{figures/vonBraun.fig12a.eps}{figures/vonBraun.fig12b.eps}
%\caption{\label{n6208_positions} This figure shows the positions of
%stars ($13.0 < I < 17.0$) in the field centered on NGC 6208 (left
%panel) and a control field (right panel) at the same Galactic latitude
%offset by 1 degree in the sky. The field is heavily dominated by field
%stars \citep{l72,pm01}, and the cluster members are not obviously
%concentrated.}
%\end{figure}

%---------------------------------------------------------------------------

%---------------------------------------------------------------------------
%---------------------------------------------------------------------------
\end{document}